\documentclass[11pt,a4paper]{article}

\usepackage[margin=0.80in]{geometry}
\usepackage{setspace}
\usepackage{amsmath,amssymb,amsfonts}
\usepackage{bm}
\usepackage{booktabs}
\usepackage{graphicx}
\usepackage{enumitem}
\usepackage[numbers,sort&compress]{natbib}
\usepackage{siunitx,booktabs}
\usepackage[hidelinks]{hyperref}
\usepackage{authblk}

\usepackage{mathpazo}

\begin{document}
	
	\title{\textbf{Modeling structure and credit risk of the economy: \\a multilayer bank-firm network approach}}
	
	\author{
		Soumen Majhi, Anna Mancini, and Giulio Cimini\\
		Physics Department and INFN, \\University of Rome Tor Vergata, 00133 Rome (Italy)
		
	}
	
	\date{\today}
	
	\maketitle

\begin{abstract}
Assessing the resilience of the economy requires accounting for its intrinsic multi-layer nature, by assessing for instance how disruptions at the firm level spread through the production network and propagate to the banking sector. Methods exist to measure the reverberation of shocks over the multilayer network of supply-customer relations among firms, corporate loans of banks and their interbank market exposures. 
However, empirical network data are often privacy protected and thus inaccessible to researchers and regulators. 
In this work we develop an unified framework, combining state-of-the art techniques to reconstruct the whole multilayer structure of the economy 
from balance sheet information of banks and firms, as well as dynamics of shock propagation from the interfirm to the interbank layers. 
We showcase application of our methodology using data of the Italian economy. We identify the most systemically important firms and industries, as well as the most vulnerable banks, further assessing the determinants of systemic risk -- obtaining results coherent with the empirical literature on network contagion. 
Overall, our framework allows performing detailed network-based stress tests on a digital twin of the economy, without requiring detailed network information that is difficult to acquire.
\end{abstract}

\section{Introduction}

Modern economies are characterized by complex multi-layered interdependencies linking corporate entities with financial institutions. Production networks describe the web of supply relationships through which firms exchange intermediate goods and services~\cite{schweitzer2009economic}, while financial networks capture credit exposures among banks and between banks and firms~\cite{bardoscia2021physics}. The structural coupling between these two domains is fundamental for the functioning of modern economies, but also plays a crucial role in determining how localized disruptions evolve into systemic crises. Assessing economic resilience therefore requires understanding not only contagion within each network layer, but also the feedback mechanisms that connect the economic and financial sectors.

Recent global events have highlighted the importance of such cross-layer interactions. Climate-related disasters~\cite{carleton2016social,carvalho2021supply,battiston2017climate}, geopolitical conflicts~\cite{goes2022impact}, and the COVID-19 pandemic~\cite{pichler2022forecasting} have revealed how disruptions in production networks can rapidly propagate across supply chains, generating firm-level losses that spill over to creditor banks. When suppliers halt production, downstream firms experience revenue declines that may translate into liquidity shortages and loan defaults, thereby impairing bank balance sheets. Financial distress can in turn feed back into the real economy through credit contraction. These reinforcing loops lie at the core of systemic risk~\cite{acemoglu2015systemic,fouque2013handbook} and economic resilience~\cite{amini2016resilience,hallegatte2014economic}. Developing tools capable of quantifying such multilayer contagion dynamics has thus become a priority for regulators and policymakers.

A fundamental obstacle to this endeavor is data availability. Detailed information on interfirm transactions, bank-firm credit exposures, and interbank lending relationships is typically confidential and inaccessible to researchers and stakeholders. Yet systemic risk is inherently shaped by network topology. Being able to reconstruct hidden economic and financial networks from partial balance-sheet information has therefore become a prerequisite for realistic stress testing and policy analysis~\cite{cimini2019statistical}.

In the aftermath of the global financial crisis of 2007/2008, research efforts concentrated primarily on the interbank market. The reconstruction problem in this context consisted in inferring bilateral exposures from aggregate balance-sheet figures. This challenge was addressed using methods rooted in statistical physics: when only aggregate information is available, the principle of maximum entropy prescribes selecting the least biased probabilistic network consistent with the available information (acting as constraint). Applied to networks, this approach leads to Exponential Random Graph (ERG) models~\cite{park2004statistical,squartini2011analytical}. Fitness-based ERG specifications have proven highly effective in reconstructing interbank networks without topological information, consistently outperforming alternative probabilistic approaches in validation exercises~\cite{anand2018missing}.
More recently, attention has shifted towards production networks. Unlike interbank markets, supply chains are shaped by technological and sectoral compatibility constraints: not all pairs of firms can realistically trade, and connections depend on input-output requirements. Standard ERG formulations, which treat all node pairs as potentially connected, are therefore insufficient to capture the functional structure of production systems. Models incorporating sectoral information and input-output constraints have therefore begun to address this limitation~\cite{ialongo2022reconstructing,fessina2024}.

Beyond reconstruction, another key limitation of the literature lies in the fragmentation of contagion modeling. A substantial body of work has examined shock propagation within financial networks~\cite{elliott2014financial,haldane2011systemic,bardoscia2015debtrank,cimini2015systemic} or within production networks~\cite{diem2022quantifying,pichler2021and}. However, most approaches treat the real economy and the financial system as largely separate domains. Only a few recent studies have attempted to link firm-level supply chains with bank exposures. For instance, Guth et al.~\cite{guth2020modeling} investigates the impact of COVID-19 supply-chain disruptions on Austria’s banking sector using aggregated input-output data. Borsos et al.~\cite{borsos2020shock} analyze bidirectional amplification between firms and banks. Tabachov{\'a} et al.~\cite{tabachova2024estimating} develop a multilayer contagion model integrating firm-level supply chains and bank-firm credit relationships, but without incorporating the interbank market. Fialkowski et al.~\cite{fialkowski2025data} studies how production shocks generate correlated defaults and spillover into interbank solvency contagion. Despite these advances, a unified data-driven framework that simultaneously reconstructs all relevant layers and models their ordered interaction remains absent.

In this work, we address this gap by developing a comprehensive computational framework that reconstructs and integrates the interbank, bank-firm, and firm-firm networks into a single multilayer economic-financial structure. Importantly, our approach relies exclusively on balance-sheet information, enabling the construction of a “digital twin” of the economy without requiring access to confidential micro-level transaction data.
Each layer is reconstructed using state-of-the-art techniques tailored to its structural properties. The interbank network is inferred using the Density-Corrected Gravity Model (DCGM)~\cite{cimini2015systemic}, the bank-firm credit layer via the Enhanced Capital Asset Pricing Model (ECAPM)~\cite{squartini2017enhanced}, and the firm-firm production network through the Stripe-Corrected Gravity Model (SCGM), using the Input-Output Gravity Model (IOGM)~\cite{ialongo2022reconstructing,fessina2024} specification. These reconstructed layers are then combined into a coherent multilayer network.

On top of this structure, we design an ordered contagion pipeline linking production and financial distress. Shocks originate in the production layer, where disruptions propagate across supply chains and generate output losses measured by the Economic Systemic Risk Index (ESRI)~\cite{diem2022quantifying}, accounting for essential and non-essential input constraints~\cite{pichler2021and}. The resulting erosion of firm profit margin translates into credit  losses for banks, quantified through the Financial Systemic Risk Index (FSRI)~\cite{tabachova2024estimating} -- which we extend to take into account non-performing loans (NPL). Residual bank-level equity losses then propagate within the interbank market via DebtRank dynamics~\cite{battiston2012debtrank}, capturing financial amplification effects.
Together, these coupled processes link the economic and financial sectors into a single cascade framework, allowing us to track how local failures evolve into systemic events.

We apply this unified framework to balance-sheet data for Italian firms and banks. Our analysis identifies the most systemically important firms and industries, as well as the most vulnerable banks, and provides empirical evidence on the determinants of systemic risk. Overall, our results demonstrate that integrating reconstructed production and financial networks within a single cascade framework enables faithful, detailed network-based stress testing of the economy, even in the absence of granular transaction data.

\begin{figure}[p]
		\centering
		\includegraphics[width=\textwidth]{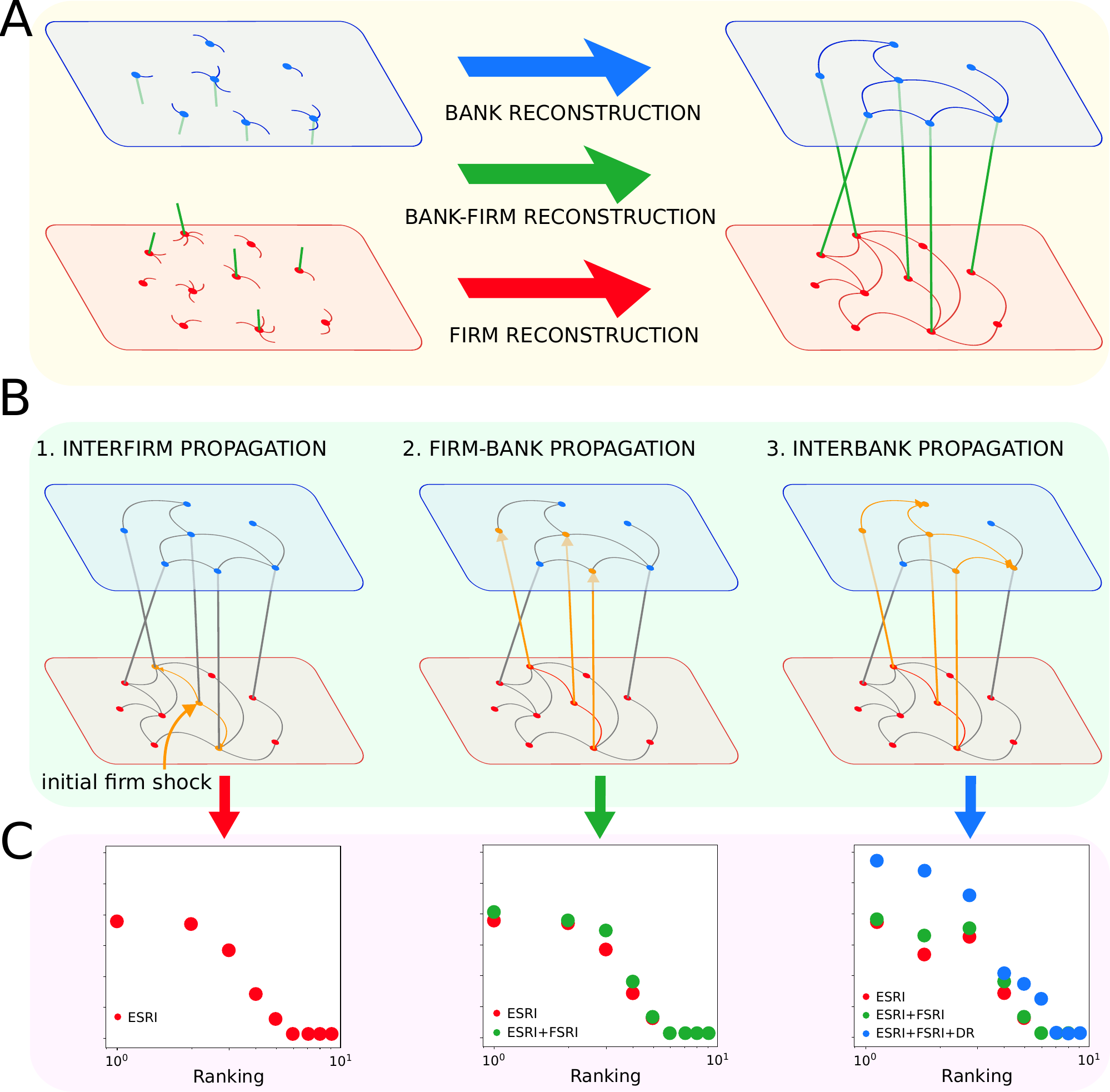}
		\caption{Schematic overview of the multilayer bank-firm systemic-risk framework. (\textbf{A}) \textit{Network reconstruction}: starting from balance-sheet information for banks and firms in our sample, three interconnected networks are reconstructed independently: the interbank market layer (top, blue links), the bank-firm credit network (middle, green links), and the interfirm production network layer (bottom, red links).
        (\textbf{B}) \textit{Shock propagation}: Once the multilayer network is reconstructed, we simulate the spreading of a shock originating in the interfirm layer. The shock first diffuses through interfirm production links, according to ESRI dynamics (left), then is transmitted to the bank layer via firm exposures to banks, according to FSRI (center), and finally amplified through interbank contagion following DR dynamics (right). (\textbf{C}) \textit{Systemic-risk analysis}: the cumulative effects of the shock propagation are summarized through ranked risk profiles, showing how successive layers contribute to economic losses: reduction of production (ESRI, left), additional bank credit losses (ESRI+FSRI, middle), and interbank amplification (ESRI+FSRI+DR, right).}
		\label{fig:schematic_framework}
	\end{figure}

Figure~\ref{fig:schematic_framework} provides a conceptual representation of the full modeling pipeline underlying our firm-bank contagion analysis.	Panel~A illustrates the reconstruction stage, where interfirm, firm-bank, and interbank layers are inferred separately and combined into a multilayer bank-firm network. Panel~B depicts the ordered propagation mechanism,where shocks originate at the firm level, spread across production relationships, induce bank distress through credit exposures, and are subsequently transmitted within the interbank market. Panel~C summarizes the resulting systemic-risk measures, emphasizing how risk rankings progressively change as additional layers of contagion are incorporated.

\section{Materials and Methods}
	
\subsection{Variables Definitions}

	We model a multi-layer weighted directed network composed of $N_B$ banks and $N_F$ firms. Banks are labeled with Greek letters ($\alpha=1,\dots,N_B$) while firms are labeled with Latin letters ($i=1,\dots,N_F$).
	Table~\ref{tab:variables} reports the value we extract from balance sheets of banks and firms, which we use to reconstruct the multilayer network (see below). 
    Bank-level attributes for each bank $\alpha$ include interbank assets $A_\alpha$, interbank liabilities $L_\alpha$, corporate loans $C_\alpha$ and equity $E_\alpha$. Firm-level attributes for each firm $i$ are production $R_i$, supply $S_i$, bank loans $B_i$, shareholder funds $E_i$ and sector of production $p_i$. 
    Finally, we consider industrial sector attributes (Input–Output table entries $\{s_{p\to p'}\}$ and essentiality relations $\{e_{p\to p'}\}$) to inform the interfirm layer through input-output flows and critical supplier dependencies (see below). 
    We exploit this balance-sheet level information as constraint to reconstruct a statistically principled ensemble of multilayer networks, which then we use as substrate for the shock propagation dynamics.
	
	 \begin{table}[h!]
		\centering
		\begin{tabular}{cl}
			\toprule
			\textbf{Variable} & \textbf{Description} \\
			\midrule
			$A_\alpha$ & Total interbank assets of bank $\alpha$ (claims on other banks) \\
			$L_\alpha$ & Total interbank liabilities of bank $\alpha$ (obligations to other banks) \\
			$C_\alpha$ & Total corporate loans of bank $\alpha$ (credit granted to firms) \\
			$E_\alpha$ & Total equity of bank $\alpha$ (net worth) \\
			\midrule
			$R_i$ & Total production/output of firm $i$ (proxied by revenues) \\
			$S_i$  & Total supply/input of firm $i$ (proxied by cost of materials and services) \\
			$B_i$      & Total bank loans of firm $i$ (borrowing from banks) \\
			$E_i$      & Total equity of firm $i$ (shareholders' funds) \\
			$p_i$      & Sector of production of firm $i$ (NACE2 classification) \\
			\midrule
			$s_{p\to p'}$ & Input–Output table entry: flow from sector $p$ to sector $p'$ (NACE2) \\
			$e_{p\to p'}$ & Essentiality relation: whether sector $p$ is an essential input for sector $p'$ (NACE2)\\
			\bottomrule
		\end{tabular}
		\caption{Variables used in our framework and their economic meaning, at the bank, firm, and industrial sector levels.}
		\label{tab:variables}
	\end{table}
	
	\subsection{Network Reconstruction}

    The following procedure, derived from constrained maximum entropy arguments augmented by a fitness ansatz \cite{cimini2019statistical}, allows defining connection probabilities and weights describing the multilayer bank-firm network. 
    Such probabilities can be used to generate individual realizations of the multilayer network, i.e., they describe an ensemble of multilayer networks. 
    The only required input is the balance sheet information described above, which is also preserved on average over the ensemble. 
    
	\subsubsection{Interbank layer $\cal B$}
	Bank-bank links define the interbank market $\cal B$. We denote the amount of the interbank loan from lender bank $\alpha$ to borrower bank $\beta$ as $w_{\alpha\to\beta}^{\cal B}$. 
    This is reconstructed through the Density-Corrected Gravity Model (DCGM)~\cite{cimini2015systemic} as:
	\begin{equation}
		w_{\alpha\to\beta}^{\cal B}=\frac{A_\alpha L_\beta}{W^{\cal B}\:p_{\alpha\to \beta}^{\cal B}}\qquad \text{with probability\:\:}p_{\alpha\to \beta}^{\cal B}=\frac{z^{\cal B}A_\alpha L_\beta}{1+z^{\cal B}A_\alpha L_\beta}
		\label{eq:DCGM}
	\end{equation}
	and $w_{\alpha\to\beta}^{\cal B}=0$ otherwise. 
    That is, the loan amount is given by the product of the interbank assets of $\alpha$, $A_\alpha$, and interbank liabilities of $\beta$, $L_\beta$, normalized by the total weight of layer $\cal B$, $W^{\cal B}$, and the connection probability $p_{\alpha\to \beta}^{\cal B}$ of the two banks, given by a Logistic function of the same quantities (known as {\em fitness model} in network science). 
    The parameter $z^{\cal B}$ sets the network density (see below), while $W^{\cal B}=\sqrt{AL}$ is the total weight of layer $\cal B$, given by the geometric mean of the two total masses $A=\sum_{\alpha}A_\alpha $ and $L=\sum_{\alpha}L_\alpha$. 
    When the system is closed (meaning that the total masses are equal, $A\equiv L$), 
	by construction the model preserves the total interbank assets and liabilities of each bank, as ensemble averages. 
    For instance, $\langle A_\alpha\rangle = \sum_\beta \langle w_{\alpha\to\beta}^{\cal B} \rangle = \sum_\beta  w_{\alpha\to\beta}^{\cal B} p_{\alpha\to\beta}^{\cal B} = A_\alpha (L/W^{\cal B}) =A_\alpha$. Deviations appear if the system is not closed: $\langle A_\alpha\rangle = A_\alpha \sqrt{L/A}$.
	
	\subsubsection{Bank-firm layer $\cal I$}
	Bank-firm links define the bipartite (inter-layer) corporate relationships $\cal I$. The amount of the loan from bank $\alpha$ to firm $i$, denoted as $w_{\alpha\to i}^{\cal I}$, is reconstructed through the bipartite version of the DCGM, also known as Enhanced Capital Asset Pricing Model (ECAPM)~\cite{squartini2017enhanced}:
	\begin{equation}
		w_{\alpha\to i}^{\cal I}=\frac{C_\alpha B_i}{W^{\cal I}\:p_{\alpha\to i}^{\cal I}}\qquad \text{with probability\:\:}p_{\alpha\to i}^{\cal I}=\frac{z^{\cal I}C_\alpha B_i}{1+z^{\cal I}C_\alpha B_i}
		\label{eq:ECAPM}
	\end{equation}
	and $w_{\alpha\to i}^{\cal I}=0$ otherwise. As for the DCGM, the loan amount is given by the normalized product of the corporate loans of $\alpha$, $C_\alpha$, and bank loans of $i$, $B_i$, while the connection probability $p_{\alpha\to i}^{\cal I}$ is a Logistic function of the same quantities. 
    As before, the parameter $z^{\cal I}$ sets the network density, while $W^{\cal I}=\sqrt{CB}$ is the total weight of layer $\cal I$, given by the geometric mean of the total corporate loans $C=\sum_{\alpha}C_\alpha$ and total bank debt $B=\sum_{i}B_i)$. 
    Again when the system is closed ($C\equiv B$), 
	by construction the model preserves the total corporate loans of each bank and total bank debt of each firm, as ensemble averages. 
    For instance, $\langle C_\alpha\rangle = \sum_i \langle w_{\alpha\to i}^{\cal I} \rangle = C_\alpha (B/W^{\cal I}) =C_\alpha$, with deviations occurring if the system is not closed, $\langle C_\alpha\rangle = C_\alpha \sqrt{B/C}$.

	\subsubsection{Firm-firm layer $\cal F$}
	Firm-firm links define the production network $\cal F$. Reconstructing this layer is more involving than the previous cases, since links are not monetary loans anymore but correspond to purchases of goods and services according to firms' production processes. 
    As customarily done in the literature \cite{ialongo2022reconstructing,diem2022quantifying}, we assume that firms produce outputs only in their own industrial sector. Therefore a link $w_{i\to j}^{\cal F}$ corresponds to the (monetary) amount of product type $p_i$ supplied by firm $i$ to customer firm $j$. 
    The Stripe-Corrected Gravity Model (SCGM)~\cite{ialongo2022reconstructing} reconstructs these links as:
	\begin{equation}
		w_{i\to j}^{\cal F}=\frac{R_i S_{p_i\to j}}{W_{p_i}^{\cal F}\:p_{i\to j}^{\cal F}}\qquad \text{with probability\:\:}p_{i\to j}^{\cal F}=\frac{z^{\cal F}R_iS_{p_i\to j}}{1+z^{\cal F}R_iS_{p_i\to j}}
		\label{eq:SCGM}
	\end{equation}
	and $w_{i\to j}^{\cal F}=0$ otherwise. 
    As before the parameter $z^{\cal F}$ sets the total network density, while now the model uses sector-specific normalization $W_{p_i}^{\cal F}=\sqrt{R_{p_i}S_{p_i}}$, 
    where $R_{p_i}=\sum_{i\in p_i}R_i$ is the total production of sector $p_i$ while $S_{p_i}=\sum_j S_{p_i\to j}$ is the total supply of firms from sector $p_i$. 
    According to the SCGM formulation, outgoing connections (sales) of firm $i$ are ruled by its total production $R_i$ (all in sector $p_i$) while, differently from DCGM, incoming connections (purchases) of firm $j$ depend on how much input its production process requires from a given sector $p$, indicated as $S_{p\to j}$. 
    Then when the system is closed for each sector ($R_{p}\equiv S_{p}$ $\forall p$), the model preserves the total production of each firm and its total input by sector, as ensemble averages: 
    $\langle R_i \rangle = \sum_j \langle w_{i\to j}^{\cal F}\rangle = R_i$ 
    and $\langle S_{p_i\to j} \rangle = \sum_{i\in p_i} \langle w_{i\to j}^{\cal F}\rangle = S_{p_i\to j}$. Deviations in this case read, for instance, $\langle S_{p_i\to j} \rangle = S_{p_i\to j} \sqrt{R_{p_i}/S_{p_i}}$. 

    However, values of the {\em costs by sector} $S_{p\to j}$ (how much $j$ buys from suppliers belonging to sector $p$) are not directly available from balance sheet information. 
    We therefore estimate them using the Input-Output Gravity Model (IOGM) approach \cite{fessina2024}:
	\begin{equation}
		S_{p_i\to j} = S_j\frac{s_{p_i\to p_j}}{\sum_{p} s_{p\to p_j}}
		\label{eq:IO_stripe}
	\end{equation}
	where $s_{p_i\to p_j}$ is the total flux from sector $p_i$ to sector $p_j$ (how much $p_j$ requires from $p_i$) and $\sum_{p}s_{p\to p_j}$ is the total incoming flux of sector $p_j$ (how much $p_j$ requires from all other sectors). These quantities are obtained from Input-Output (IO) tables~\cite{Miller:2009aa,leontief1986input}. In other words, we split the total costs of firm $j$ across industrial sectors proportionally to what IO tables prescribe.

	\subsection{Shock Propagation and Systemic Risk Metrics}

Given a reconstructed network sampled according to the procedure described above, we simulate contagion spreading from the production network layer to the interbank network layer according to the three steps described in detail below:
	\begin{enumerate}
		\item Quantify the amount of disruption in the production network after an initial shock, using ESRI dynamics \cite{diem2022quantifying};
		\item Update the banks' equity buffers due to non-performing loans to disrupted firms, generalizing FSRI dynamics \cite{tabachova2024estimating};
		\item Propagate credit shocks in the interbank market due to non-performing loans of affected banks, using DR dynamics \cite{bardoscia2015debtrank}.
	\end{enumerate}

	\subsubsection{Shocks in the production network: the Economic Systemic Risk Index}
	
	The Economic Systemic Risk Index (ESRI)~\cite{diem2022quantifying} quantifies the output reduction experienced by the whole production network after an initial production shock hitting one or more firms. 
	The initial shock is described by a vector $\psi$, with generic component $\psi_j$ representing the remaining production level of firm $j$. 
	For instance, $\psi_j=0$ and $\psi_i=1$ $\forall i\neq j$ represents the initial default of firm $j$, while other firms remain unaffected. 
Shock propagation is then ruled by a generalized Leontief production function: inputs from essential sectors ($p\in\text{Ess}_i$) set a hard constraint on the output of firm $i$, while the inputs from non-essential sectors ($p\notin\text{Ess}_i$) are treated in a linear way. 
The distinction between essential and non-essential inputs is derived from~\cite{pichler2021and,pichler2020production}, hence for each pair of sectors $p$ and $p'$ we have the sector essentiality relation $e_{p\to p'}=1$ if $p$ is an essential input for $p'$, while $e_{p\to p'}=0$ otherwise. 
We thus have $\text{Ess}_i=\{p:e_{p\to p_i}=1\}$.

ESRI dynamics is based on the downstream impact matrix, whose element $\Lambda_{ji}^\text{d}$ determines the fraction of production firm $i$ loses if firm $j$ stops supplying to it:
\begin{equation}  \label{eq:algorithm_lambda_d}
	\Lambda_{ji}^\text{d} = \begin{cases}
		\frac{w_{j\to i}^{\cal F}}{S_{p_j\to i}}\qquad \text{if} \; p_j \in \text{Ess}_i \\
		\frac{w_{j\to i}^{\cal F}}{S_i}\qquad \text{if} \; p_j \notin \text{Ess}_i
	\end{cases}
\end{equation}
and the upstream impact matrix, whose element $\Lambda^\text{u}_{ji}$ determines the fraction of production firm $i$ loses if firm $j$ stops buying from it:
\begin{eqnarray} \label{eq:lambda_u}
	\Lambda^\text{u}_{ji} = \frac{w_{i\to j}^{\cal F}}{R_i}
\end{eqnarray}

After applying the initial shock, downstream and upstream shocks are propagated to any firm $i$ through two iterative equations:

\begin{equation} \label{eq:esri_downstream}
	h_i^{\text{d}}(t+1)   =   \min\Big[
	\min_{p \in \text{Ess}_i} \Big(\tilde{\Pi}_{ip}(t)   \Big), \;
	\tilde{\Pi}_{i}(t)
	, \psi_i \Big]
\end{equation}
\begin{equation} \label{eq:esri_upstream}
	h^u_{i}(t+1) = \min \Big[ \sum_j \Lambda^\text{u}_{ji}h^\text{u}_j(t), \psi_i  \Big]  
\end{equation}
Here $t$ indicates the time step of the propagation, while $h_{i}^{d}(t)$ and $h_{i}^{u}(t)$ are the residual fraction of production of firm $i$ at time $t$ following the propagation of the downstream and upstream shock, respectively, with initial values $h_i^d(0)=h_i^u(0)=\psi_i$. 
The relative amount of essential inputs in sector $p\in\text{Ess}_i$ available for firm $i$ is:
\begin{equation}
\tilde{\Pi}_{ip}(t) = 1 - \sum_{j\in p} \sigma_j(t) \Lambda^{d}_{ji} \big(1-h_j^\text{d}(t) \big)
\end{equation}
while the relative share of all non-essential inputs is: 
\begin{equation}
\tilde{\Pi}_{i}(t) = 1 - \sum_{p\notin \text{Ess}_i} \sum_{j\in p}^n \sigma_j(t) \Lambda^{d}_{ji} \big(1-h_j^\text{d}(t) \big)
\end{equation}
and firms' market share, used as a proxy for how replaceable a firm is for its buyers, is:
\begin{equation}
\sigma_j(t) =  \min \left( \frac{R_j}{\sum_{l\in p_j} R_l \,h_l^d(t)},1 \right)
\label{eq.mshare}
\end{equation}

    After the two, independent shocks of eqs. \eqref{eq:esri_downstream} and \eqref{eq:esri_upstream} have converged at $t=t^*$, the residual fraction of output of firm $i$ is computed as 
	\begin{equation}
		h_i(t^*)=\min\{h_i^{d}(t^*),h_i^{u}(t^*)\}
	\end{equation}
	The ESRI value is then given by the fraction of total production lost in the system:
	\begin{equation}
		\text{ESRI}(\psi)=\sum_{i}\frac{R_i}{\sum_jR_j}[1 - h_i(t^*)]
	\end{equation}

	\subsubsection{Credit shocks to the bank layer: the Financial Systemic Risk Index}

	The reduced production levels, $h(t^*)$, correspond to a drop in firms' revenues and material costs that affects their profits. Reduced profits in turn affect the equity of firms and, eventually, the ability to repay their bank loans -- which become non-performing loans (NPL). 
	Indeed the production reduction of firm $i$ translates into a reduction of profit as: 
	\begin{equation}
		\Delta \pi_i = \max\left\{0,\Big[1-h_i^u(t^*) \Big]R_i - \left[1-h_i^d(t^*) \right]S_i\right\}
	\end{equation}
	where $R_i$ is the revenue and $S_i$ the material costs, which are reduced proportionally to the upstream and downstream shocks, respectively 
    \footnote{With this formula we link the reduction in production costs to the downstream shock, i.e. the supply shock, and the reduction in revenues to the upstream, or demand shock. We also take only positive values of $\Delta \pi$. Overall, this is different from the formula $\Delta \pi_i = \left( 1-h_i(t^*) \right)(R_i - S_i)$ used in \cite{tabachova2024estimating}. Additionally, we do not consider the effect of reduced profit on liquidity buffers and potential insolvency of firms.}.
    This causes a direct reduction of firm's equity: $E_i' = \max[0,\,E_i - \Delta \pi_i]$. 
	We then assume that the value of the loan $w_{\alpha\to i}^{\cal I}$ that bank $\alpha$ has towards firm $i$ reduces proportionally to the equity loss of $i$:
	$\left(w_{\alpha\to i}^{\cal I}\right)' = w_{\alpha\to i}^{\cal I}\left(E_i'/E_i\right)$.
	Hence when $i$ has no profit reduction ($\Delta \pi_i=0$) then loans keep their original value, while when the equity of $i$ becomes 0 (meaning that the firm has defaulted), the entire exposure is written off.
	Overall, considering losses from all firms, the reduced equity of bank $\alpha$ is calculated as $E_\alpha ' = \max\left\{0,
	E_\alpha - \sum_i [w_{\alpha\to i}^{\cal I}-(w_{\alpha\to i}^{\cal I})']\right\}$,
	so the relative equity loss for bank $\alpha$ is
	\begin{equation}
		h_\alpha(1)=1-\frac{E_\alpha '}{E_\alpha}=\min\left\{1,\frac{1}{E_\alpha}\sum_i w_{\alpha\to i}^{\cal I}\left(1-\frac{E_i'}{E_i}\right)\right\}=\min\left\{1,\sum_i \frac{w_{\alpha\to i}^{\cal I}}{E_\alpha}\min\left(1,\:\frac{\Delta \pi_i}{E_i}\right)\right\}
		\label{h1bank}
	\end{equation}
    Differently from \cite{tabachova2024estimating}, we do not only count banks that default, but quantify the reduction of banks equity due to production shocks of firms.
    The overall loss for the banking system is the Financial Systemic Risk Index \cite{tabachova2024estimating} 
	\begin{equation}
		\text{FSRI}(\psi)=\sum_{\alpha}\frac{E_\alpha}{\sum_\beta E_\beta}h_\alpha(1)
		\label{FSRIdef}
	\end{equation}
	which is the equity-weighted sum of losses suffered by individual banks in the network, and represents the fraction of total bank equity that is lost after the initial shock propagates through the supply chain network, generating NPLs.

	\subsubsection{Credit shocks in the interbank market: the DebtRank}
	
	Finally we use the Debt Rank algorithm (DR)~\cite{battiston2012debtrank,bardoscia2015debtrank} to propagate credit shocks originating from bank losses $\{h(1)\}$ in the interbank layer. The idea at the basis of DR is that credit shocks propagate also in the absence of defaults, provided that balance sheets are deteriorated: potential losses in the equity of a borrower translate into the devaluation of interbank assets of the corresponding lender. 
	Losses are then obtained by iteratively spreading the individual banks distress levels weighted by the potential wealth affected. 
	
	The level of financial distress 
	of each bank $\alpha$ at each time step $t$ of the shock propagation dynamics is given by the relative change of equity with respect to the original value: $h_\alpha(t)=1-E_\alpha(t)/E_\alpha$. 
	By definition, $h_\alpha=0$ when no equity losses occurred for bank $\alpha$, $h_\alpha=1$ when that bank defaults, and $0<h_\alpha<1$ for intermediate distress levels. 
	In our case the initial shock comes from NPL, thus
	$h_\alpha(1)=1-E_\alpha'/E_\alpha$ as of eq. \eqref{h1bank}.
	Subsequent values of $h$ are obtained by spreading this shock on the interbank layer, 
	according to the equation describing the evolution of banks equity. 
	By defining $\mathcal{A}(t)=\{\beta:h_\beta(t-1)<1\}$ as the set of banks that have not defaulted up to time $t$ (and thus can still spread their financial distress), 
	we assume that a generic bank $\beta$ propagates shocks as long as it keeps receiving them, i.e., provided that $h_\beta(t)>h_\beta(t-1)$. 
	We have:
	\begin{equation}
    \label{eq:acca}
		h_\alpha(t+1)=\min\left\{1,\; h_\alpha(t)+\sum_{\beta\in\mathcal{A}(t)}\frac{w_{\alpha\to\beta}^{\cal B}}{E_\alpha}\,[H_\beta(t)-H_\beta(t-1)]\right\} 
	\end{equation}
    where $H_\beta(t)$ is the probability of default of bank $\beta$ at time $t$. 
    As in \cite{bardoscia2016distress}, we assume the form 
    \begin{equation}
    H_\beta(t)=h_\beta(t)e^{\eta [h_\beta(t)-1]}
    \end{equation}
    where $\eta\ge0$ is a free parameter representing the inverse of the typical relative equity loss after which banks start to propagate distress to their creditors. $\eta$ allows interpolating between two of the mostly widely used contagion models: $\eta=0$ leads to the linear DebtRank (where the default probability is the relative equity loss), while $\eta\to\infty$ recovers contagion-at-default (or Furfine algorithm, where the probability of default is one only if the equity is depleted, and zero otherwise).

	The described dynamics stops at $t^*$ when no more banks can propagate their distress, i.e., $H_\alpha(t^*)=H_\alpha(t^*-1)$ $\forall \alpha$. 
	The overall DebtRank of the network is then obtained as:
	\begin{equation}
        \label{eq:drank}
		DR(\psi)=\sum_\alpha\frac{E_\alpha}{\sum_\beta E_\beta}h_\alpha(t^*)
	\end{equation}
	$DR$ thus represents the amount of equity that is potentially at risk in the system, given an initial shock $\{h_\alpha(1)\}$. 
    Note that while in our setup DebtRank is a function of the initial shock at the firm level, $\psi$, we can compute a resilience indicator for individual banks by averaging over different realizations of the initial shock. 
    We thus define the {\em vulnerability} of bank $\alpha$ as \cite{cimini2016entangling}:
    \begin{equation}
        \label{eq:vulnerability}
		V_\alpha=\langle h_\alpha(t^*|\psi)\rangle_\psi    
    \end{equation}
    where $h_\alpha(t^*|\psi)$ specifies the final value of $h$ obtained by using $\psi$ as initial condition of the dynamics, and the average is performed over all possible default configurations $\psi_i$ of the chosen setup (see below).

	\section{Data Description}

    \subsection{Dataset Construction}

    We inform the multilayer framework using a proprietary dataset covering Italian banks and firms for the year 2023, compiled from regulatory balance sheets. 
    
    Bank data is obtained from the BankFocus database by Bureau van Dijk. We applied 4 filtering steps to select the relevant banks for our analysis: 1) Status: active company; 2) World region: Italy; 3) Specialization: commercial bank, finance company, cooperative bank, savings bank, specialized governmental credit institution, private banking, investment bank; 4) Total assets in 2023 $> 2\,000\,000$€, with the exclusion of companies with no recent financial data and Public authorities. The resulting bank dataset is composed of $N_B=109$ Italian banks.
    
    Firm data is obtained through AIDA, Bureau van Dijk's database of Italian firms. We applied the following filters: 1) Companies who are ultimate owners of their groups; 2) Companies whose NACE2 code is not 64, 65, 66 (financial companies are removed); 3) Total assets for 2023 $\geq 10\,000$€ and total shareholder's funds for 2023 $\geq 1\,000$€. The resulting firm data is composed of $N_F=7109$ Italian firms.

    Input-output (IO) table entries $\{\sigma_{p\to p'}\}$ are obtained from ISTAT (Italian Institute of Statistics)\footnote{\url{https://www.istat.it/tavole-di-dati/il-sistema-di-tavole-input-output-anni-2020-2022/}} for the last available year 2022. 
    Given that we only have a subsample of the Italian economy, we rescale IO entries such that the total output of a sector matches the total revenues of firms in that sector for our data. This means that the factors entering eq. \eqref{eq:IO_stripe} are computed as: 
\begin{equation}
s_{p\to p'}=R_p \frac{\sigma_{p\to p'}}{\sum_{p''}\sigma_{p\to p''}}
\end{equation}

	\subsection{Descriptive Statistics}

\begin{figure}[h!]
		\centering
		\includegraphics[width=\textwidth]{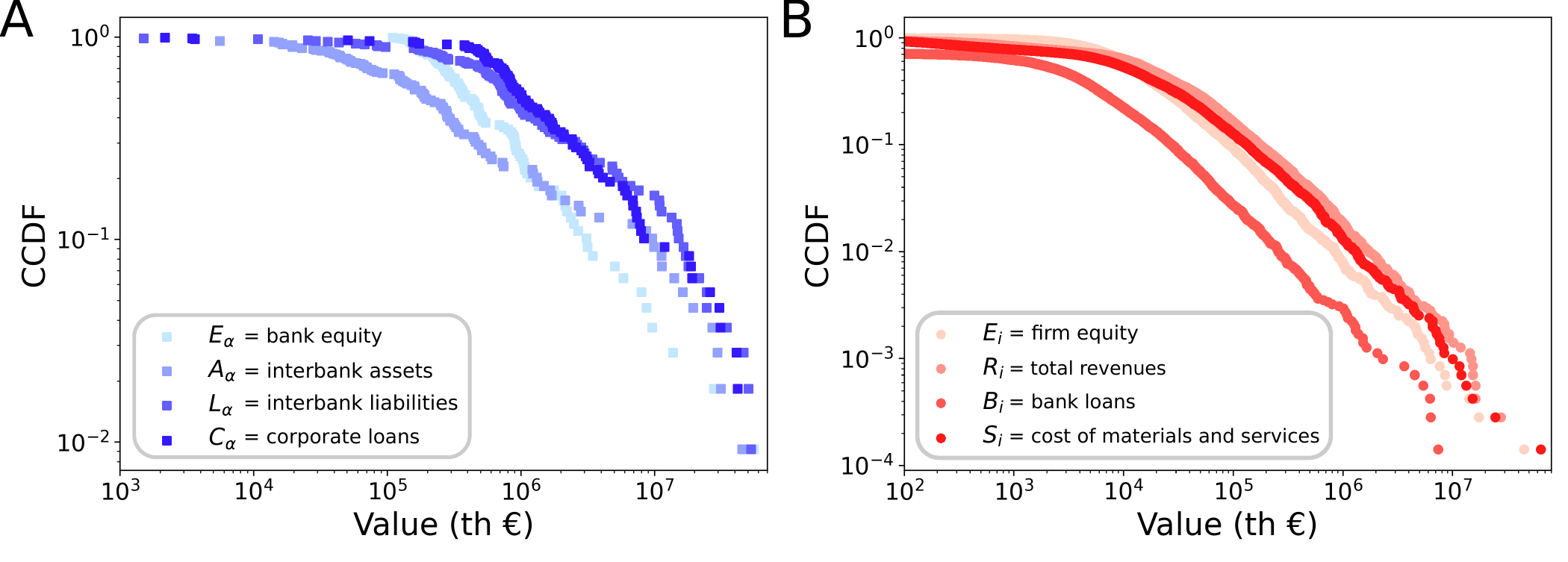}
        \caption{CCDF of bank variables (A) and firm variables (B) extracted from the balance sheets of our 2023 sample of the Italian economy.}
		\label{fig:stats}
	\end{figure}

    Tables \ref{tab:bankstats} and \ref{tab:firmsstats} in the Appendix report summary statistics (mean, standard deviation, minimum and maximum, skewness) for the balance-sheet variables of our sample that we use to build the multilayer framework, while Figure \ref{fig:stats} reports the corresponding complementary cumulative distribution functions (CCDF). 
    For all quantities, the distribution shows a power law tail, highlighting the wide heterogeneity of banks and firms: most institutions have intermediate balance sheet figures while a few very large ones dominate the upper tail. This is confirmed by mean values being much higher than median ones, with maximum values being orders of magnitude larger than typical ones. These patterns may suggest that systemic importance is concentrated in a small subset of large banks and firms, with shocks to top-ranked firms potentially dominating aggregate outcomes. 
    Therefore in the following we will perform sensitivity analyses to examine tail dependence and robustness with respect to the upper end of the distributions.
    Note that all CCDFs saturate at relatively high values, since our sample covers only the most capitalized banks and firms in Italy. The only exception is bank debt of firms, which is 0 for about 20\% of firms in our sample.

    \section{Results: Reconstructed networks}
    
    \subsection{Density setting and reconstructed degrees}
    
    Before running the multilayer network reconstruction framework, we need to specify the three free parameters, $z^{\cal B}$, $z^{\cal I}$ and $z^{\cal F}$, which determine the link density of the interbank, bank-firm and interfirm layers, respectively. 
    As the level of connectivity strongly influences systemic risk properties of the network \cite{ramadiah2020network}, we want to set these values properly using empirical measures provided by the literature.
    However we must notice that, in real networks, the link density $\rho$ typically scales with the inverse of the number of nodes, $\rho\sim N^{-1}$. This means that we cannot directly use density values taken from empirical studies about networks of different sizes, especially because our network only represents a subset of the Italian economy. 
    A more reliable quantity is the average node degree $\langle k\rangle \sim \rho N$, which in principle does not vary with the size of the network. Indeed, in our case it represents the intrinsic capacity of a bank or firm to connect with others, which should not depend on the size of the economy. 
    We thus proceed as follows.
    
    In the interbank layer, according to eq. \eqref{eq:DCGM} each bank $\alpha$ has an average out-degree (number of borrower banks) equal to $\langle k_\alpha^{out,\cal B}\rangle = \sum_{\beta} p_{\alpha\to \beta}^{\cal B}$ and and average in-degree (number of lending banks) equal to $\langle k_\alpha^{in,\cal B}\rangle = \sum_{\beta} p_{\beta\to \alpha}^{\cal B}$. We thus set $z^{\cal B}$ so that the mean values of these quantities over banks equals 20 \cite{manna2009topology,finger2013network}, 
    corresponding to a density value of $\rho_B=0.17$:    
    \begin{equation}
z^{\cal B}\mbox{ such that }\frac{1}{N_B}\sum_\alpha\langle k_\alpha^{out,\cal B}\rangle\equiv \frac{1}{N_B}\sum_\alpha\langle k_\alpha^{in,\cal B}\rangle=\frac{1}{N_B}\sum_{\alpha\neq\beta} p_{\alpha\to \beta}^{\cal B}=20
    \end{equation}

    Similarly in the interfirm layer, 
    according to eq. \eqref{eq:SCGM} each firm $i$ has an average out-degree (number of customer firms) equal to $\langle k_i^{out,\cal F}\rangle = \sum_j p_{i\to j}^{\cal F}$ and and average in-degree (number of supplier firms) equal to $\langle k_i^{in,\cal F}\rangle = \sum_j p_{j\to i}^{\cal F}$. We set $z^{\cal F}$ so that the mean values of these quantities over firms equals 40 \cite{bacilieri2022firm}, corresponding to a density value of $\rho_F=0.0054$: 
\begin{equation}
z^{\cal F}\mbox{ such that }\frac{1}{N_F}\sum_i\langle k_i^{out,\cal F}\rangle\equiv \frac{1}{N_F}\sum_i\langle k_i^{in,\cal F}\rangle=\frac{1}{N_F}\sum_{i\neq j} p_{i\to j}^{\cal F}=40
\end{equation}
    
    For the bank-firm layer, according to eq. \eqref{eq:ECAPM} each bank $\alpha$ has an average out-degree (number of borrower firms) equal to $\langle k_\alpha^{out,\cal I}\rangle = \sum_i p_{\alpha\to i}^{\cal I}$ while each firm has an average in-degree (number of lending banks) equal to $\langle k_i^{in,\cal I}\rangle = \sum_\alpha p_{\alpha\to i}^{\cal I}$. 
    We set $z^{\cal I}$ so that the average firm degree equals 1.8 \cite{demasi2012bankfirm}, corresponding to an average bank degree of 117.5 and a density value of $\rho_F=0.0.016$: 
\begin{equation}
z^{\cal I}\mbox{ such that }\frac{1}{N_F}\sum_i\langle k_i^{in,\cal I}\rangle=\frac{1}{N_F}\sum_{i,\alpha} p_{i\to j}^{\cal I}=1.8 \Longrightarrow \frac{1}{N_B}\sum_\alpha\langle k_\alpha^{out,\cal I}\rangle=117.5
\end{equation}

\begin{figure}[p]
	\centering
	\includegraphics[width=\textwidth]{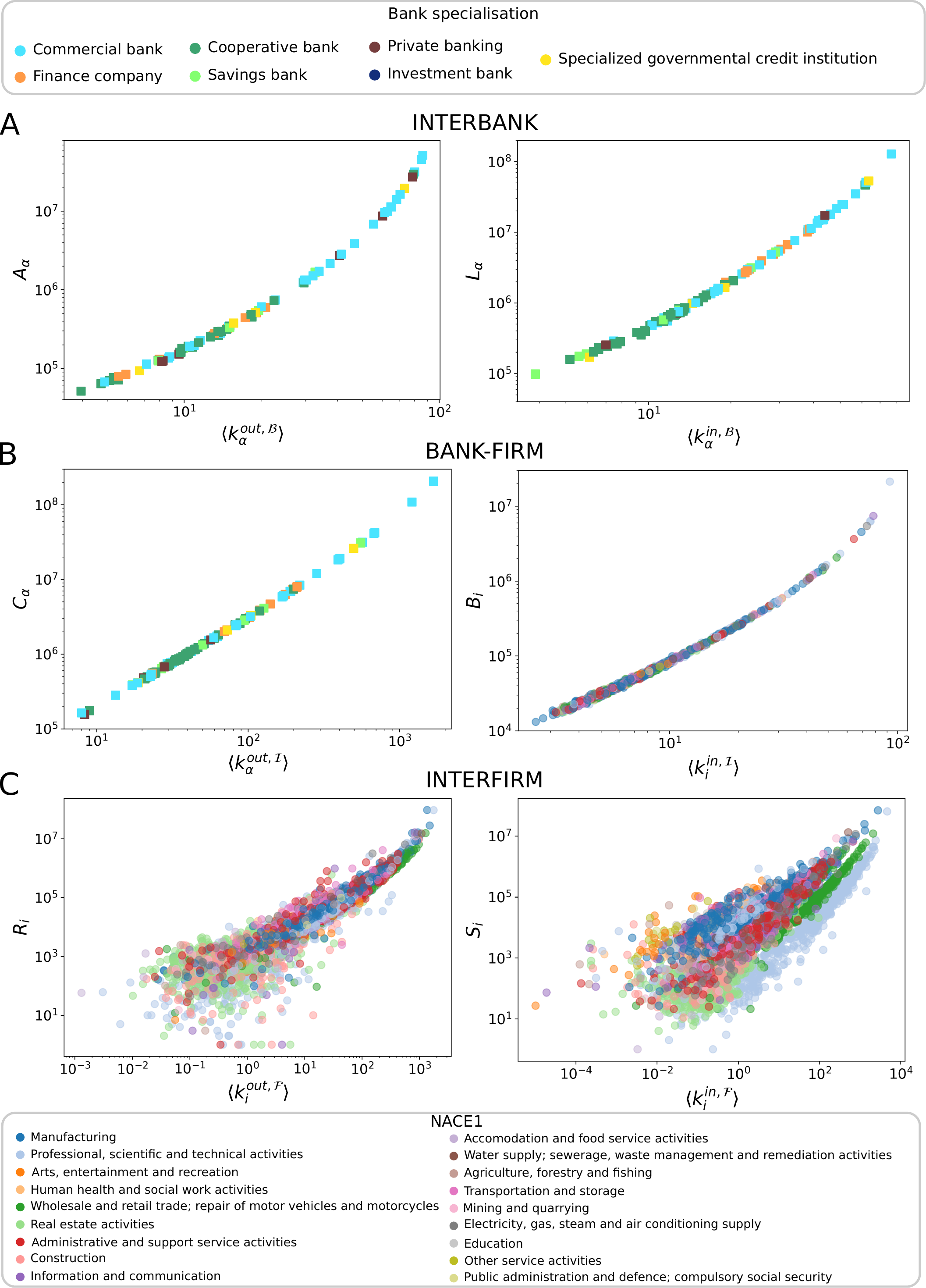}
    \caption{Scatter plots of node degrees versus strength used in the reconstruction procedure. Each point represents an individual bank or firm, colored according to their specialization or NACE1 sector, respectively.}
	\label{fig:reconstructed_deg}
\end{figure}

    Using these values, we can generate independent samples of the multilayer network. 
    Results presented below are computed as average values over an ensemble of 100 networks.
    Figure \ref{fig:reconstructed_deg} shows how the reconstruction procedure determines node degrees. Each value is scattered versus the balance sheet quantity that mainly determines it. For instance, the average out-degree of a bank in the interbank layer reads $\langle k_\alpha^{out,\cal B}\rangle = \sum_{\beta} p_{\alpha\to \beta}^{\cal B}\sim A_\alpha$ at first approximation when $A_\alpha$ is small. Analogously, $\langle k_\alpha^{in,\cal B}\rangle \sim L_\alpha$, 
    $\langle k_\alpha^{out,\cal I}\rangle \sim C_\alpha$, $\langle k_i^{in,\cal I}\rangle \sim B_i$, $\langle k_i^{out,\cal F}\rangle \sim R_i$, $\langle k_i^{in,\cal F}\rangle \sim S_i$. Naturally, these linear relationships saturate for large balance sheet values, as connection probabilities cannot grow above one and so degrees are upper-bounded by the number of nodes in the network. Overall, this makes the degree distribution highly skewed with an upper cutoff. Note that for $k_i^{out,\cal F}$ and $k_i^{in,\cal F}$ we observe slightly different trends for the various industrial sectors, which depend on their level of balance (how much $R_{p}$ and $S_{p}$ differ) in a non-linear way.

	\subsection{Strength conservation}

\begin{figure}[p]
	\centering
    \includegraphics[width=\textwidth]{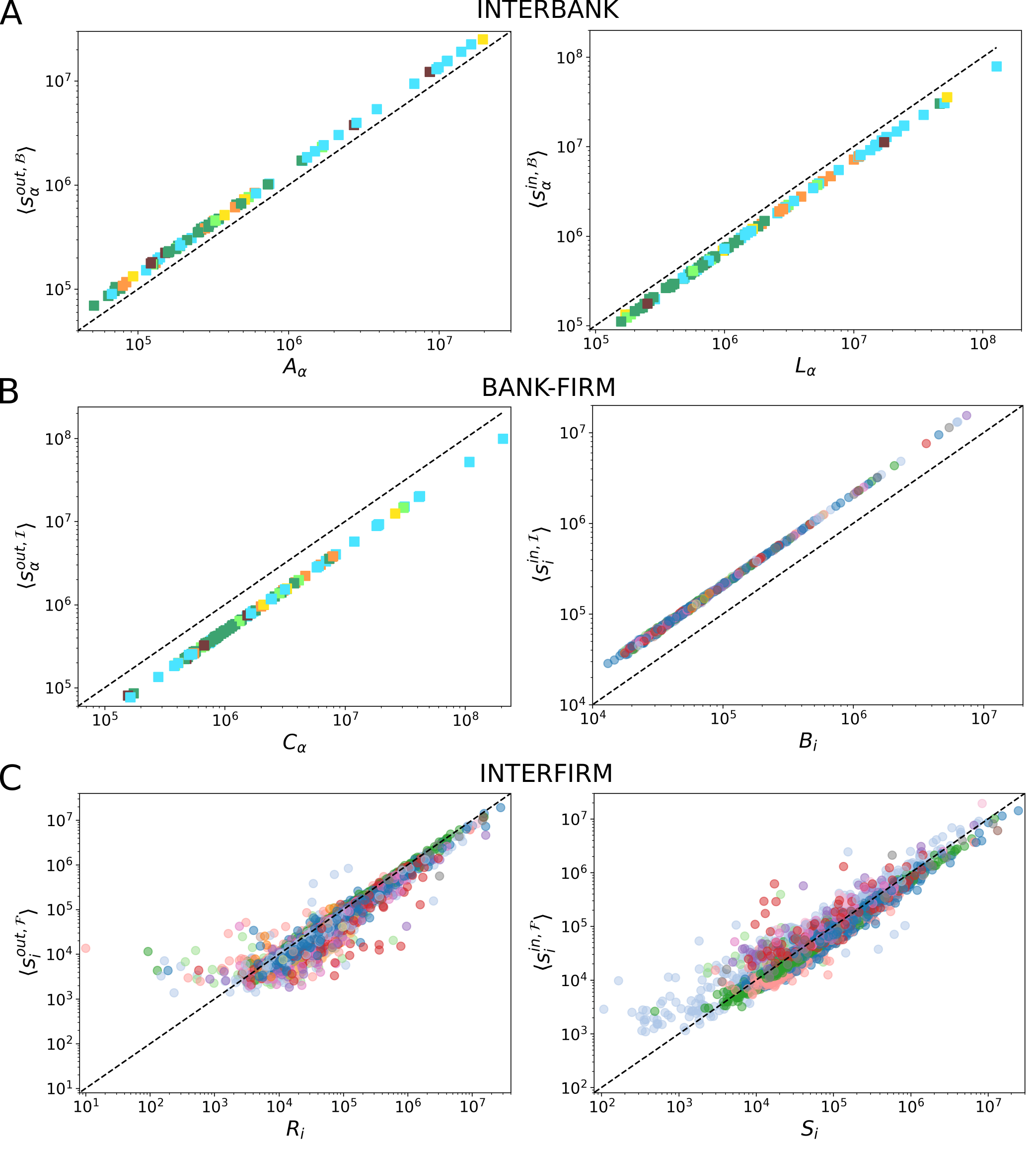}
	
    \caption{Scatter plots of empirical versus reconstructed strength values, where each point represents an individual bank or firm, while the identity (solid line) serves as a reference for perfect reconstruction. Banks are colored according to their specialization while firms to their NACE1 sector.}
	\label{fig:reconstructed_str}
\end{figure}

    Here, we check whether the reconstruction process is able (as intended) to reproduce balance sheet variables of banks and firms -- which we generically refer to as {\em strengths}. This validation step is crucial, as it confirms that the reconstructed network retains the key balance-sheet figures necessary for subsequent contagion and systemic-risk dynamics.
	Figure~\ref{fig:reconstructed_str} shows scatter plots of empirical versus reconstructed strength values. Most points, representing individual banks or firms, lay along the identity line, indicating that the method reproduces empirical strengths with high accuracy. Offsets are due to the system not being closed (for instance, reconstructed interbank assets/liabilities are systematically higher/lower than their empirical counterparts), but such deviations are generally small. 

\newpage

\section{Results: Systemic risk of individual firms}

We now discuss simulation results of the shock propagation dynamics on the multilayer network, studying the impact of the failure of individual firms. We thus represent the default of firm $j$ with the initial condition $\psi_j=0$ and $\psi_i=1$ $\forall i\neq j$ (all other firms are unaffected). 
Given such initial shock, we run ESRI dynamics of eqs. \eqref{eq:esri_downstream} and \eqref{eq:esri_upstream} up to convergence at $t^*$, then do FSRI dynamics of eq. \eqref{h1bank} and finally run DR dynamics eq. \eqref{eq:acca} up to convergence at $t^*$, using the nonlinear formulation with $\eta=2$ (we avoid using linear propagation otherwise the banking layer suffers overwhelming unrealistic losses).
With this scenario, we measure the overall losses that the failure of a firm would cause to the whole economy, without any mitigation action taken on the system. 

\subsection{Ranking plots}

\begin{figure}[p]
    \centering
    \includegraphics[width=\textwidth]{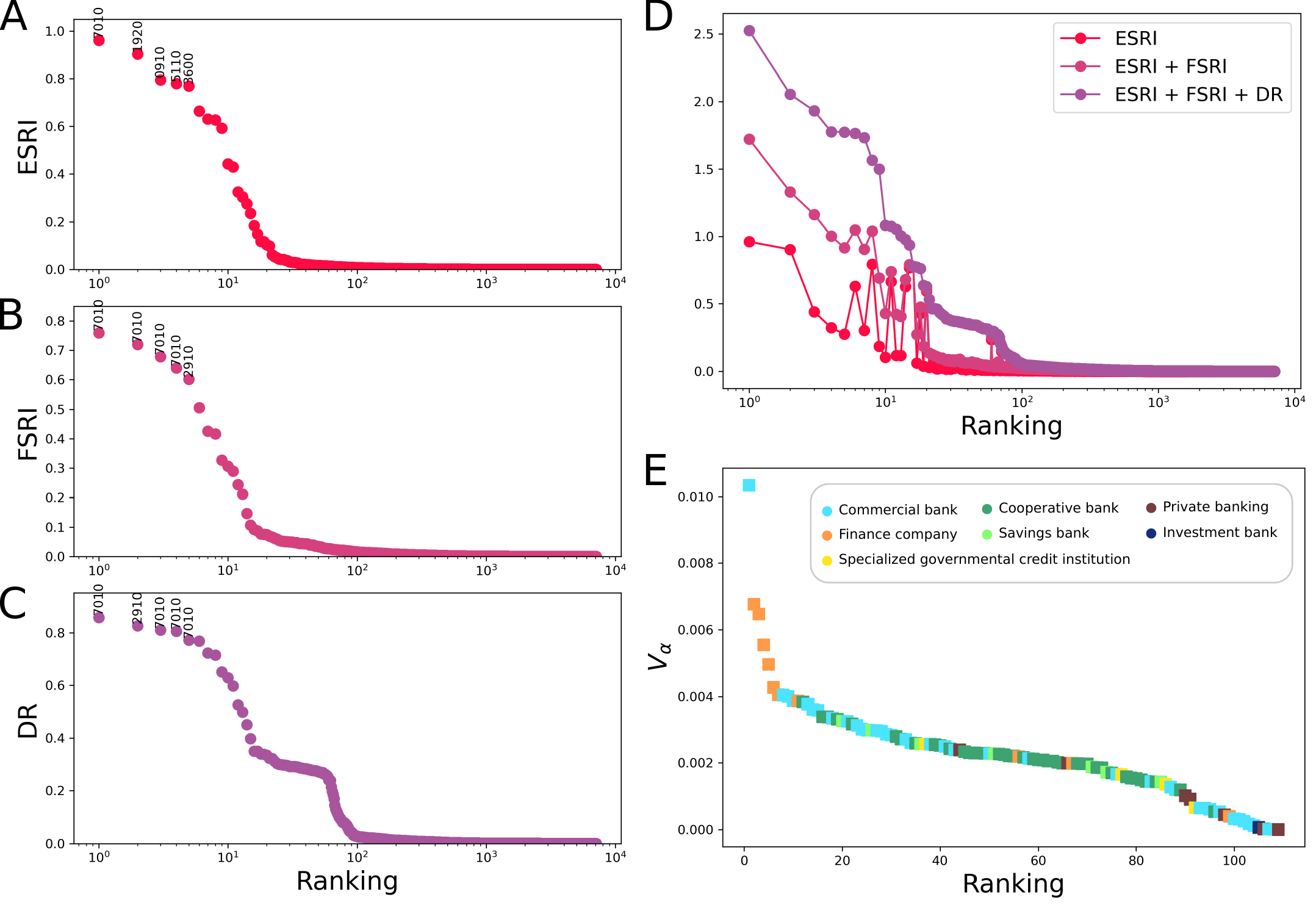}

   \caption{Ranking plots in terms of systemic risk metrics, with nodes placed in descending order of the corresponding scores. 
   (A) Ranking based on ESRI scores of firms, where for each firm $i$ the initial condition is given by its complete production shutdown ($\psi_i=0.0$), while all other firms remain active. ESRI of $i$ thus measures the fraction of total production lost in the interfirm layer through supply-chain contagion, due to the default of $i$. 
   (B) Ranking based on FSRI scores of firms: FSRI of $i$ measures the fraction of total bank equity lost in the bank-firm layer due to the results of the ESRI dynamics with the same initial condition as before.
   (C) Ranking based on DR scores of firms: DR of $i$ measures the fraction of total bank equity lost in the interbank layer due to the results of the FSRI dynamics with the same initial condition as before.
   (D) Ranking based on total systemic risk scores of firms (ESRI+FSRI+DR), highlighting the total impact of firms default on the overall economy. 
   Total impact of $i$ thus measures the fraction of total firm production and bank equity lost due to the default of $i$, subsequent supply chain contagion, bank loans devaluation and re-evaluation of interbank claims.
   (E) Ranking based on vulnerability scores of banks, obtained by averaging over shutdowns of all individual firms as initial conditions. Vulnerability of $\alpha$ thus measures how much the bank can be affected by economic losses, in terms of equity losses. Marker colors denote bank specialisation.}
    \label{fig:firm_ranking_conv}
\end{figure}

Figure~\ref{fig:firm_ranking_conv}A shows the ranking of individual firms in terms of their ESRI values (the fraction of total production lost in the interfirm network after their initial default). 
Similarly to what was found in related studies on empirical production networks \cite{diem2022quantifying,fessina2024,mancini2025evolution}, the ranking curve features an initial region of a handful high-impact firms, with ESRI values close to 1, meaning that their initial collapse can ultimately destroy the entire production of the economy (we recall that we are considering only the most capitalized firms in Italy). The top-5 firms belong to NACE4 sectors 7010 (Business management), 1920 (Manufacture of refined petroleum products), 0910 (Petroleum and natural gas extraction services), 5110 (Air passenger transport) and 3600 (Water production, treatment and supply).
The ranking curve then declines quickly: only about 0.3\% of the firms have ESRI larger than 0.1, while the rest of firms have smaller or negligible impact, indicating a strong heterogeneity of the ESRI distribution.

Figure~\ref{fig:firm_ranking_conv}B displays firms' ranking in terms of their FSRI values (the fraction of total bank equity lost in the bank-firm network after their initial default). Again we observe an initial region of high-impact firms, with FSRI up to 0.8. The top-4 firms in this case belong to NACE4 sector 7010 (Business management), while the 5th to NACE4 2910 (Manufacture of motor vehicles). FSRI then quickly declines to negligible values: again only about 0.3\% of firms have FSRI larger than 0.1. 
Figure~\ref{fig:firm_ranking_conv}C finally shows firms' ranking in terms of their DR values (the fraction of total bank equity lost in the interbank network after their initial default). The initial region of high-impact firms, with DR around 0.8, basically overlaps with that of FSRI. However in this case we see a slower decline towards negligible values, characterized by an additional flat region of medium-impact firms (DR around 0.3). 
Indeed, in this case more than 1\% of firms have DR larger than 0.1. This signals the presence of many firms whose failure has little impact on the economy but a considerable one on the financial system. 

As the three rankings just discussed mostly do not coincide (for instance a firm that ranks top according to ESRI may be in the DR rank tail), in order to assess the overall impact of firms on the multilayer system we rank them according to the sum of their ESRI, FSRI and DR scores. As Figure~\ref{fig:firm_ranking_conv}D shows, 
top-ranked ESRI firms can have very different impact on the banking sector. While FSRI is close to zero for firms with NACE4 category 3600 and 5110, the other systemically risky firms achieve top overall impact, along with those firms having intermediate ESRI values but large corporate loans, which allow their shock to travel and propagate within the banking sector.

At last, concerning banks' vulnerability, the rank plot of Figure~\ref{fig:firm_ranking_conv}E shows that banks are hardly immune to shocks originating in the interfirm layer. The distribution is not particularly heterogeneous, with most values laying in the range 0.1\%-0.4\%. Due to their specialization, finance companies typically have the largest values, while the most vulnerable one is a commercial bank (which was placed under extraordinary administration in 2023 after inspections by Bank of Italy).
Other commercial banks, as well as cooperative and savings banks tend to populate the middle of the ranking, while investment and specialized institutions are more prevalent in the lower tail.

\subsection{Empirical analysis of determinants}


After assessing the outcomes of systemic risk dynamics, we now aim at understanding the drivers of the various impact metrics. 

\subsubsection{ESRI}

Important quantities that enter in the computation of the ESRI of firm $i$ are its revenues $R_i$, production costs $S_i$,  market share $\sigma_i$ of eq. \eqref{eq.mshare}, industrial sector $p_i$ and essentiality relations with other sectors $\{e_{p_i \to p}\}$, as well as the topology of the production network around it. 
We thus estimate a minimal cross-sectional specification for ESRI:
\begin{equation}
\log(\mathrm{ESRI}_i) = \gamma
+ \beta_1 \log(\Theta_i)
+ \beta_2 \sigma_i
+ \beta_3 \mathcal{E}_i
+ \varepsilon_i
\label{eq:reg_ESRI}
\end{equation}
where $\Theta_i=\max[R_i,S_i]$ measures firm size, while $\mathcal{E}_i$ is the ``essentiality'' score of the firm in the economy, defined as
\begin{equation}
    \mathcal{E}_i=\frac{\sum_j w_{i\to j}^{\cal F}\,R_j\,e_{p_i\to p_j}}{\sum_j w_{i\to j}^{\cal F}\,R_j}
    \label{eq.essentiality}
\end{equation}
This index measures the share of downstream economic activity supplied by firm $i$ that relies on the essential inputs the firm provides. Alternatively, it can be seen as a weighted average of essentiality indicators across customers, where weights reflect the economic importance of the relationship. Hence, the essentiality index is closely related to Bonacich centrality in production networks \cite{bonacich}.

\begin{figure}[h]
	\centering
    \includegraphics[width=\textwidth]{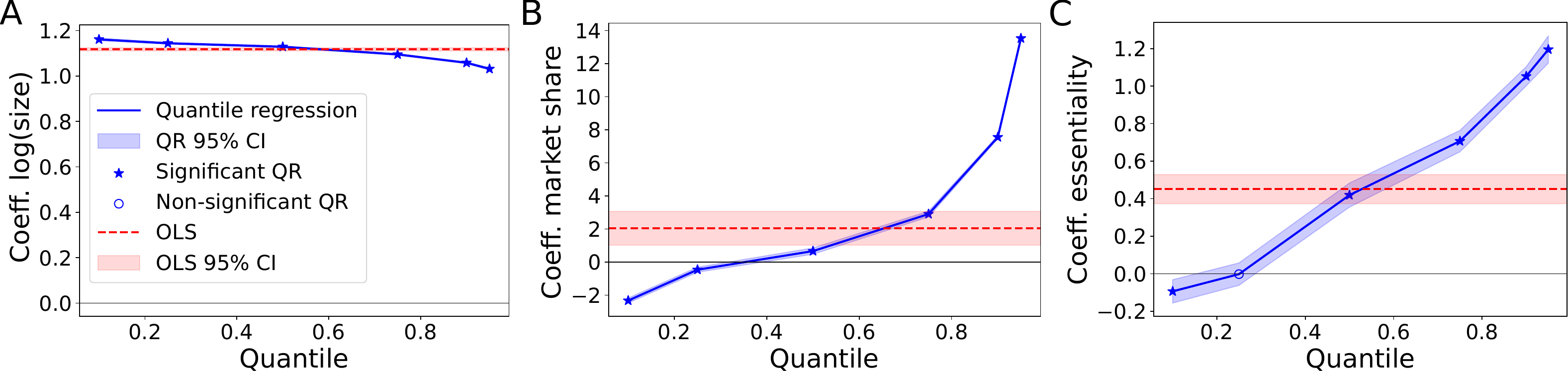}
	\caption{Determinants of ESRI for firm-level shocks: Size effects and significance of firm size (A), market share (B) and essentiality score (C) in OLS and quantile regressions of ESRI values.}
\label{fig:firm_reg_ESRI_conv}
\end{figure}

We estimate model \eqref{eq:reg_ESRI} both by OLS with heteroskedasticity-robust (HC3) standard errors and by quantile regression to explore heterogeneity across the ESRI distribution. 
Results, reported in Figure \ref{fig:firm_reg_ESRI_conv}, indicate a significant association of ESRI with these variables -- the OLS estimates show an extremely high explanatory power ($R^2=0.979$). Firm size has a strong and stable positive effect: the coefficient slightly above one implies that systemic risk increases more than proportionally with firm size. Essentiality and market share also have a positive and highly significant effect in the OLS specification, indicating that firms providing essential inputs and occupying a larger position in their market can affect a greater portion of economic activity.

Quantile regression reveals substantial heterogeneity across the ESRI distribution. While the effect of firm size remains positive and relatively stable across all quantiles, the impact of essentiality and market share increases markedly toward the upper tail. In particular, essentiality is negative or insignificant in the lower quantiles but becomes strongly positive from the median upward, reaching large and highly significant values in the top deciles. A similar pattern is observed for market share, whose effect grows dramatically at higher quantiles of ESRI. In other words, while size is a pervasive driver of systemic exposure, the combination of essential inputs and large market positions acts as a powerful nonlinear amplifier in the upper tail, shaping the concentration of extreme systemic risk.

\subsubsection{FSRI}

\begin{figure}[h]
	\centering
	\includegraphics[width=\textwidth]{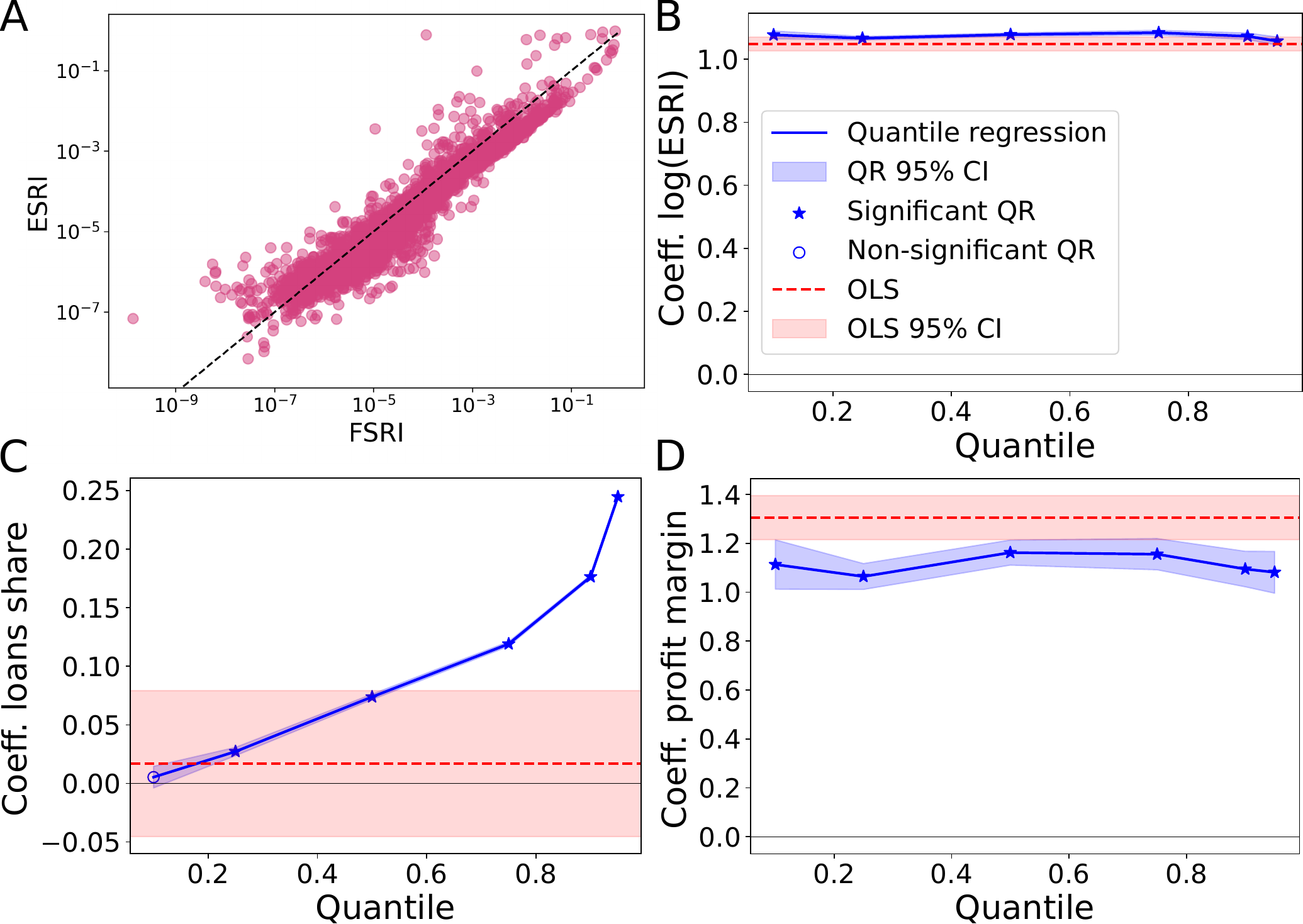}
	\caption{Determinants of FSRI for firm-level shocks.  Correlation between ESRI and FSRI (A). Size effects and significance of ESRI (B), bank loans (C) and profit margin (D) in OLS and quantile regressions of FSRI values.}
\label{fig:firm_reg_FSRI_conv}
\end{figure}

Relevant variables involved in the computation of the FSRI of firm $i$ are its size $\Theta_i$, bank loans $B_i$, and the initial shock -- proportional to ESRI$_i$ and depending also on profit margin $\mu_i=\max(R_i-S_i,0)$. 
Again we expect firm size to be an important determinant of systemic risk. According to the previous analysis, we take ESRI as a proxy of size and observe a very high correlation between FSRI and ESRI values (Figure \ref{fig:firm_reg_FSRI_conv}A); however, firms with large ESRI can differ by orders of magnitude in FSRI, indicating that financial losses transmitted to banks do depend also on firms exposures. 
We therefore estimate the following minimal cross-sectional specification:
\begin{equation}
\log(\mathrm{FSRI}_i) = \gamma
+ \beta_1 \log(\mathrm{ESRI}_i)
+ \beta_2 (B_i/\Theta_i)
+ \beta_3 (\mu_i/\Theta_i)
+ \varepsilon_i
\label{eq:reg_FSRI}
\end{equation}
where we take ESRI as a proxy of both size and initial shock, while the other two variables are normalized to avoid collinearity with size.
Results, reported in Figure \ref{fig:firm_reg_FSRI_conv}, 
reveal a strong relationship between FSRI and ESRI. In the OLS specification, with $R^2=0.935$, $\log$(ESRI) exhibits a coefficient close to one (1.048) and is highly statistically significant, indicating an almost proportional relationship between the two systemic risk measures: firms that are more systemically relevant in the production network also tend to be more relevant from a financial perspective. The coefficient of profit margin is positive and strongly significant: more profitable firms are more sensitive to the initial shock and tend to exhibit higher financial systemic relevance. By contrast, loans share is not statistically significant in the mean regression. 

The quantile regression results provide a more nuanced picture by allowing the relationship to vary across the distribution of FSRI. The coefficients of $\log$(ESRI) and profit margin remain remarkably stable while the effect of loans share increases sharply across quantiles: it is insignificant at the 10th percentile, modestly positive around the median, and becomes very large and highly significant in the upper tail of the FSRI distribution. Therefore, bank credit exposure becomes particularly relevant in determining the firms that are highly systemically important.

\begin{figure}[h!]
	\centering
    \includegraphics[width=\textwidth]{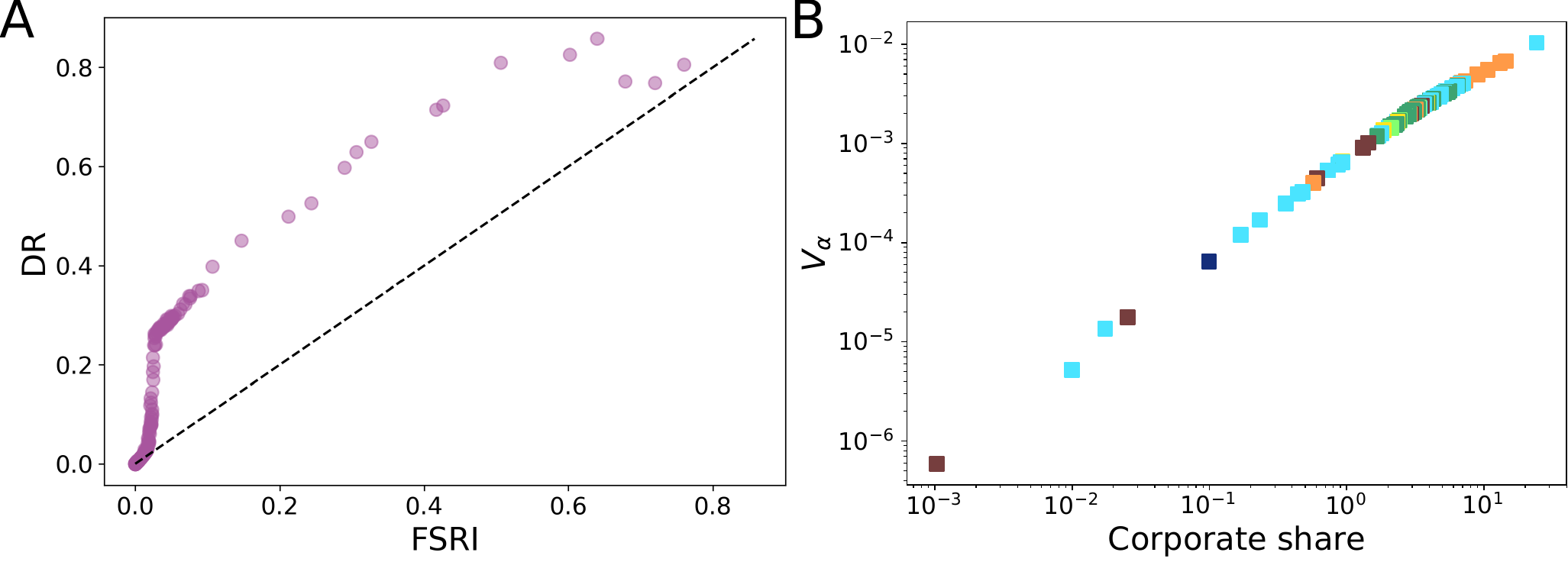}
	\caption{Determinants of DebtRank and bank vulnerability for firm-level shocks. Correlation between DR and FSRI (A) and between vulnerability and corporate loans share (B).}
\label{fig:firm_reg_DR_conv}
\end{figure}

\subsubsection{DebtRank and Vulnerability}

The situation for the DebtRank case is more involving, as the dynamics is tied to properties of the interbank layer, while here we are considering how firm characteristics trigger and drive DR. 
However, as Figure \ref{fig:firm_reg_DR_conv} shows, DR is directly driven by FSRI, with saturation effects as DR approaches the value of one. 
Results of the OLS regression of $\log(\mathrm{DR})$ versus $\log(\mathrm{FSRI})$ show an almost one-to-one relationship between FSRI and DebtRank ($R^2=0.997$): FSRI alone explains virtually all cross-sectional variation in DebtRank.  

Vulnerability instead can be more easily mapped to bank variables. However, as Figure \ref{fig:firm_reg_DR_conv} shows, there is almost perfect correlation between $V_\alpha$ and $C_\alpha/E_\alpha$, namely the proportion of corporate loans over the bank's equity. 
This confirms that corporate loan share alone almost entirely explains the cross-sectional variation in vulnerability ($R^2=0.986$).

\section{Results: Systemic risk of industrial sectors}

We now study the impact of losses stemming from all firms within an industrial sector. Therefore we use as initial condition a 10\% production loss for each firm belonging to sector $p$, identified at the NACE2 level: 
$\psi_j=0.9$ if $p_j=p$ (firm $j$ belongs to sector $p$) and $\psi_i=1$ otherwise (all other firms are unaffected).
Differently from the previous scenario, here we consider a one-step propagation dynamics: we run ESRI for one step ($t=1$), then do FSRI contagion and finally run DR for one step ($t=1$), using the linear formulation $\eta=0$. Hence, in this scenario we study the short-term impact of the initial shock, i.e., the losses that occur before any mitigation action can possibly take place. 

Figure \ref{fig:sector_rank} in the Appendix shows the basic balance sheet characteristics of each sector $p$: 
total size, defined by summing the sizes of firms belonging to it: $\Theta_p=\sum_{i\in p}\Theta_i$; 
essentiality score, defined by averaging the essentiality index of its firms: 
$\mathcal{E}_p=\langle \mathcal{E}_i\rangle_{i\in p}$; 
total bank loans $B_p=\sum_{i\in p}B_i$ and
total margin $\mu_p=\sum_{i\in p}\mu_i$.

\subsection{Ranking plots}

\begin{figure}[h!]
    \centering
    \includegraphics[width=0.93\linewidth]{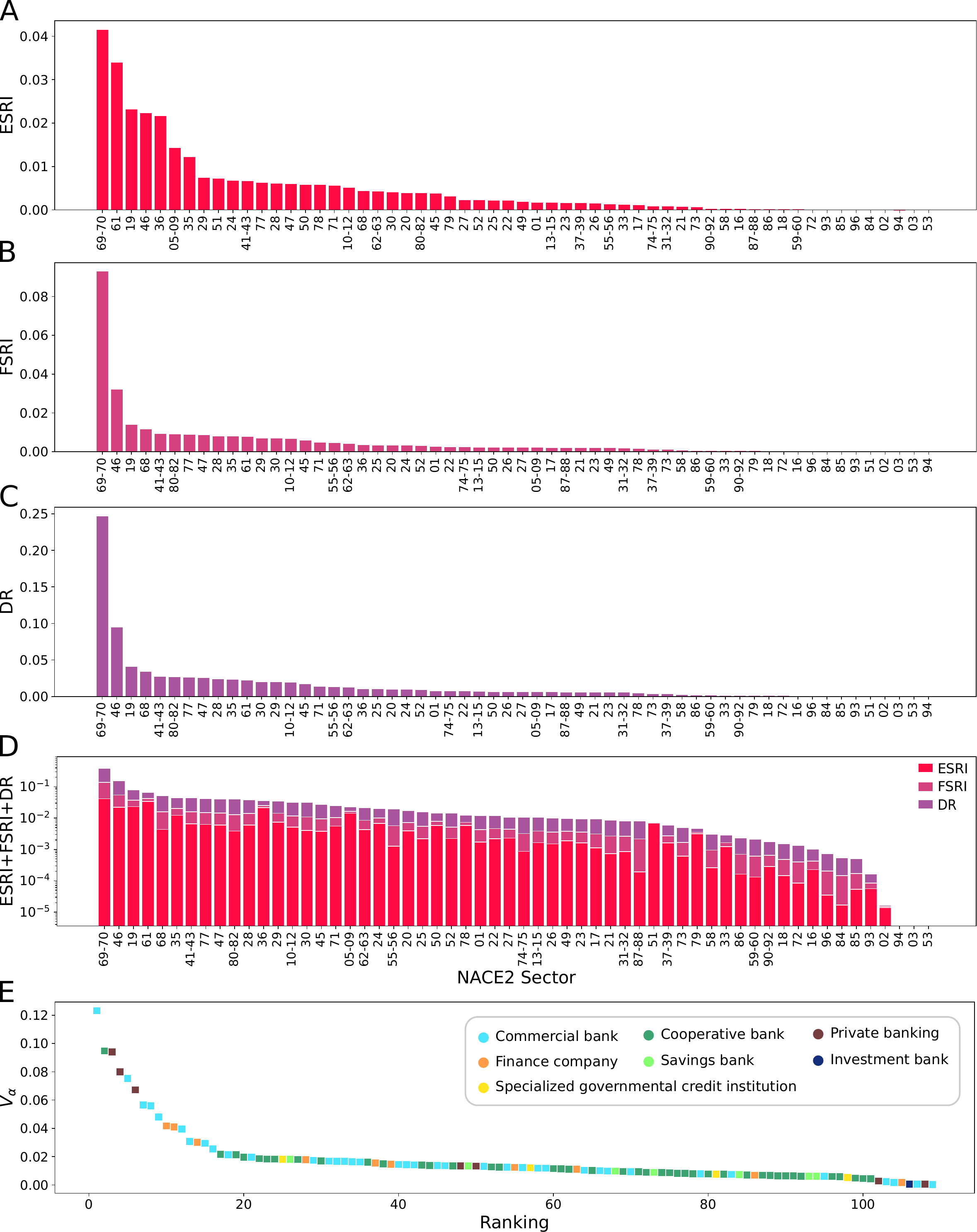}
   \caption{Ranking plots in terms of systemic risk metrics, with nodes placed in descending order of the corresponding scores. 
   (A) Ranking based on ESRI scores of sectors, where for each sector the initial condition is given by a 10\% reduction of production, while all other sectors remain fully active. 
   (B) Ranking based on FSRI scores of sectors.
   (C) Ranking based on DR scores of sectors.
   (D) Ranking based on total systemic risk scores of sectors (ESRI+FSRI+DR), quantifying the total impact of 10\% production reduction of the sector on the overall economy. 
   Each bar is decomposed into its economic (ESRI), bank-level (FSRI), and interbank (DebtRank) components, highlighting the relative contribution of each layer to aggregate systemic importance. 
   (E) Ranking based on vulnerability scores of banks, obtained by averaging over 10\% shocks of each sectors firms as initial conditions. Marker colors denote bank specialisation.}
\label{fig:sect_ranking_conv}
\end{figure}

Figure~\ref{fig:sect_ranking_conv} shows the sector rankings in terms of systemic risk scores, with sectors identified by their NACE Rev.~2 two-digit codes. 
Since in each sector we now aggregate firms of different sizes, rankings show less heterogeneity than in the case of individual firms. 
The top ESRI value of 0.04 belongs to sectors 
69–70 (Legal, accounting, management consultancy and head offices), which is the biggest and most essential overall. Other top sectors, whose decrease of 10\% productivity would cause at least a 1\% contraction of the whole economy, are:
61 (Telecommunications), 
19 (Manufacture of coke and refined petroleum products), 
46 (Wholesale trade, except of motor vehicles and motorcycles), 
36 (Water collection, treatment and supply), 
05–09 (Mining and quarrying support and extraction activities), 
35
(Electricity, gas, steam and air conditioning supply). 
At the lower tail of the ranking we find sectors 
94 (Activities of membership organisations), 03 (Fishing and aquaculture) and 53 (Postal and courier activities -- we only have one firm belonging to this sector in our data).

The FSRI and DR rankings reveal that sector 69–70 has an even larger impact on the financial system, due to its high bank loans and profit margin. Indeed, a 10\% production contraction in this sector would cause a loss of 25\% of the total bank equity. 
Other high-impact sectors are 
46 (Wholesale trade, except of motor vehicles and motorcycles), 
19 (Manufacture of coke and refined petroleum products), 
68 (Real estate activities) and 
41 (Construction of buildings), but roughly 25\% of sectors have DR larger than 0.02.

The stacked ranking plot shows the relative contribution of ESRI, FSRI and DR for each sector (we use logarithmic scale here to highlight the contribution of the different metrics also for all sectors and not just the top-impact ones). With respect to the previous scenario of individual firm defaults, now the initial shock is more broadly distributed, causing network amplification effects especially in the interbank layer (where propagation is now linear). 
As a result, bank vulnerability values are about 10 times higher, with most banks above 0.01 and the most vulnerable ones around 0.1 (they lose on average 10\% of equity for a 10\% production reduction in any NACE2 sector). 
\newpage
\newpage
\subsection{Empirical analysis of determinants}

\subsubsection{ESRI}

\begin{figure}[h]
	\centering
    \includegraphics[width=\textwidth]{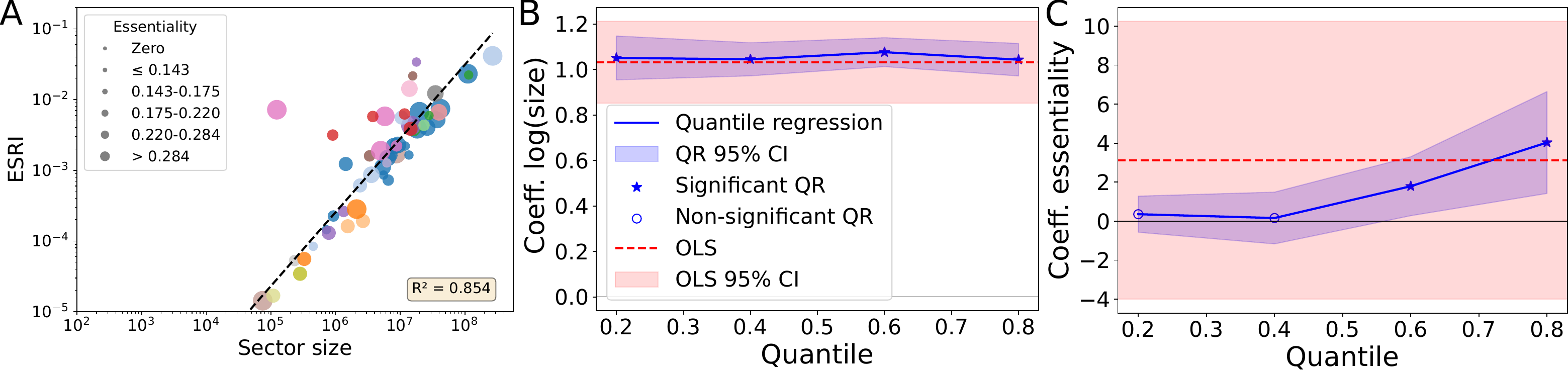}
	\caption{Determinants of ESRI for sector-level shocks. Correlation between ESRI and sector size (A). Size effects and significance of sector size (B) and sector essentiality (C) in OLS and quantile regressions of ESRI values.}
\label{fig:sect_reg_ESRI_1st}
\end{figure}

Repeating the same exercise as above, we estimate a minimal cross-sectional specification for ESRI of sectors:
\begin{equation}
\log(\mathrm{ESRI}_p) = \gamma
+ \beta_1 \log(\Theta_p)
+ \beta_2 \mathcal{E}_p
+ \varepsilon_p
\label{eq:reg_ESRI_2}
\end{equation}
(now we cannot consider the market share of a sector, which is one by definition). 
Regression results (shown in Figure \ref{fig:sect_reg_ESRI_1st}) indicate that sector size is again the primary determinant of ESRI, while the role of sector essentiality appears heterogeneous across the distribution of systemic risk. In the OLS specification, the model explains a large share of the variation in ESRI ($R^2=0.87$); the coefficient of $\log(\Theta)$ is close to one and highly significant, while the essentiality index is not: being an essential sector does not systematically increase ESRI once size is controlled for. In the quantile regressions, the elasticity of ESRI with respect to size remains remarkably stable and strongly significant, while the coefficient of essentiality becomes positive and statistically significant in the upper quantiles (0.6 and especially 0.8). Therefore, essential sectors do not necessarily have high impact, except when they are very large.

\subsubsection{FSRI}

\begin{figure}[h]
	\centering
	\includegraphics[width=\textwidth]{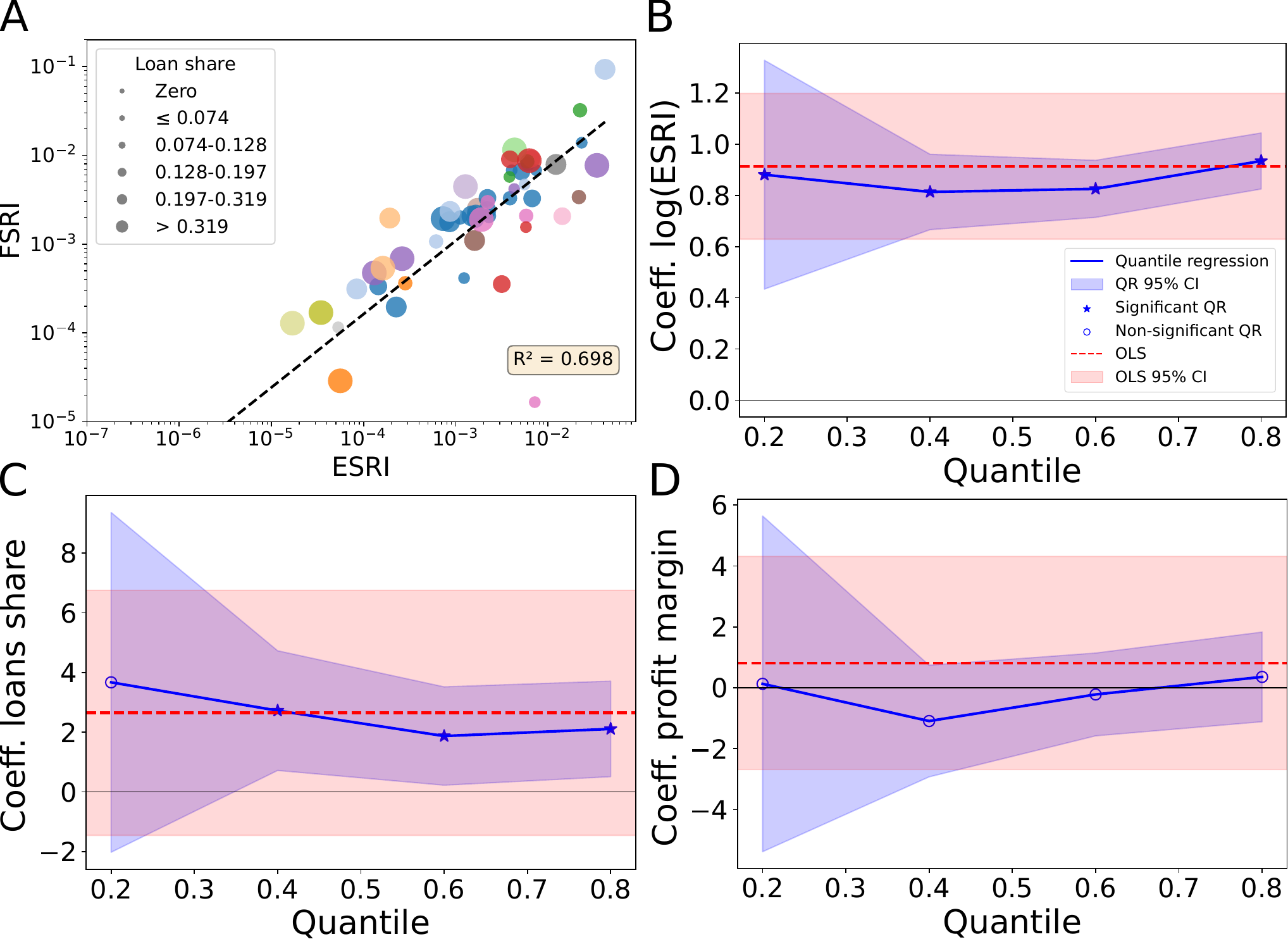}\\
    \caption{Determinants of FSRI for sector-level shocks. Correlation between FSRI and ESRI (A). Size effects and significance of ESRI (B), bank loans share (C) and profit margin (D) in OLS and quantile regressions of FSRI values.}
\label{fig:sect_reg_FSRI_1it}
\end{figure}

Considering FSRI, Figure \ref{fig:sect_reg_FSRI_1it}A shows the positive correlation with ESRI: sectors that generate large economic losses tend to be more financially damaging, but the mapping is far from one-to-one -- due to differences in bank credit exposures. 
We therefore estimate the following minimal cross-sectional specification:
\begin{equation}
\log(\mathrm{FSRI}_p) = \gamma
+ \beta_1 \log(\mathrm{ESRI}_p)
+ \beta_2 (B_p/\Theta_p)
+ \beta_3 (\mu_p/\Theta_p)
+ \varepsilon_i
\label{eq:reg_FSRI}
\end{equation}
Regression results, reported in Figure \ref{fig:sect_reg_FSRI_1it}, indicate that financial systemic risk is strongly driven by economic systemic relevance, with balance sheet characteristics playing a more limited, heterogeneous role. In the OLS specification, the model explains a substantial fraction of the variation in FSRI ($R^2 = 0.73$). The coefficient of ESRI is large, positive, and highly significant, with an elasticity close to unity. By contrast, loans share $(B/\Theta)$ and profit margin $(\mu/\Theta)$ are not statistically significant, meaning that, on average, financial exposure and profitability do not systematically affect FSRI once real systemic importance is accounted for.

In the quantile regressions, the coefficient of ESRI remains highly significant and remarkably stable across all quantiles. In contrast, loans share becomes positive and statistically significant from the median quantile onward (0.4, 0.6, and 0.8), indicating that leverage through bank financing contributes to systemic exposure mainly for firms that are already structurally relevant in the economy. Finally, profit margin is never statistically significant and its coefficients fluctuate in sign across quantiles, providing little evidence that firm profitability systematically affects financial systemic importance. Overall, these findings suggest that financial systemic risk largely reflects the economic counterpart, with credit exposures acting as a reinforcing mechanism for firms that are already systemically important.

\subsubsection{DebtRank and Vulnerability}

\begin{figure}[h]
	\centering
    \includegraphics[width=\textwidth]{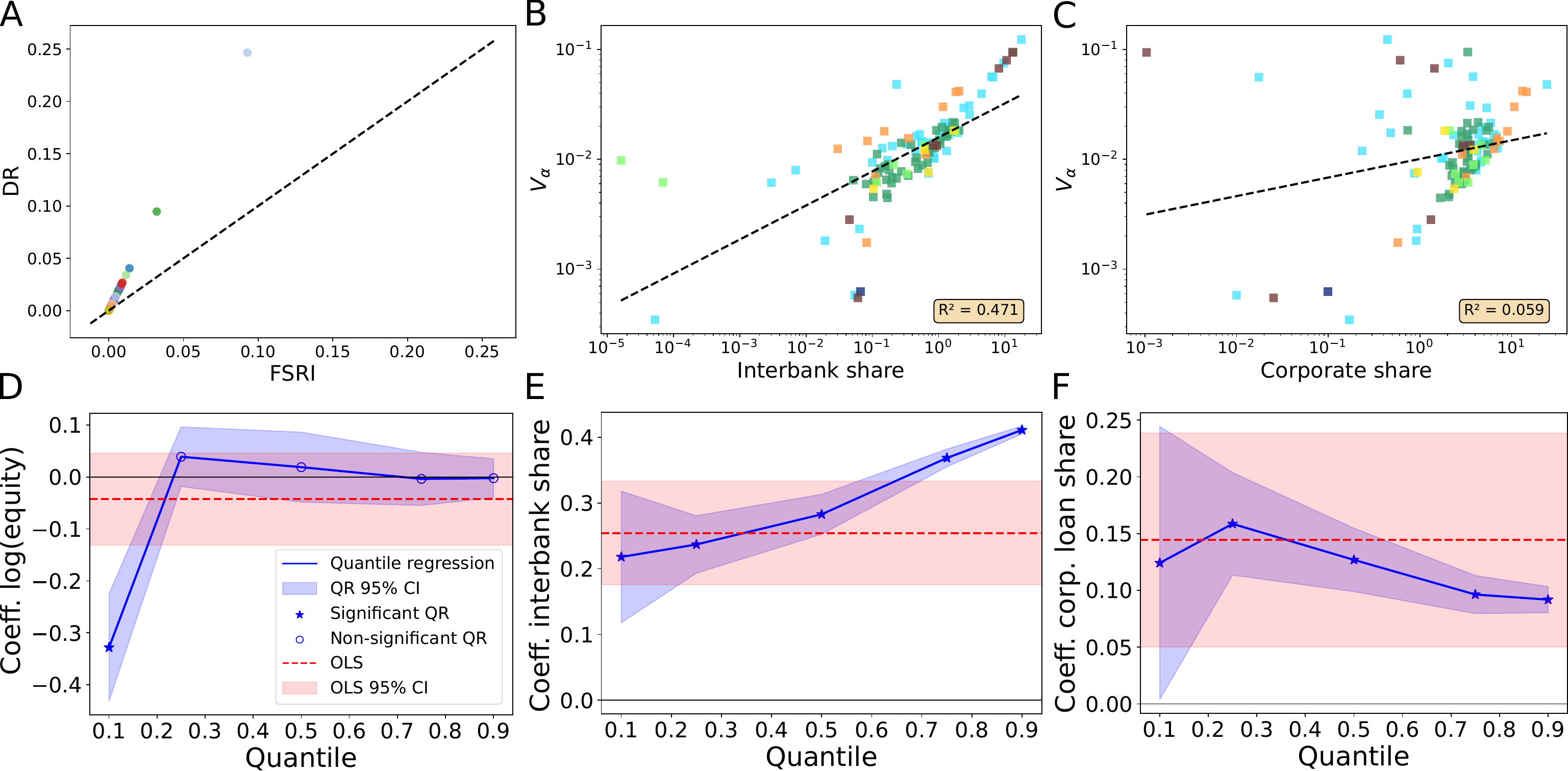}
	\caption{Determinants of DebtRank and bank vulnerability for sector-level shocks. Correlation between DR and FSRI (A), between vulnerability and interbank share (B), between vulnerability and corporate loans share (C). 
    Size effects and significance of equity (D), interbank loans share (E) and corporate loans share (F) in OLS and quantile regressions of vulnerability values.}
\label{fig:sect_reg_DR_1it}
\end{figure}

Figure \ref{fig:sect_reg_DR_1it}A 
shows that DR is directly driven by FSRI ($R^2=1.000$), 
implying that sectoral DebtRank is nearly a deterministic, proportional amplification of the initial bank-equity losses captured by FSRI. Indeed, FSRI represents the initial condition of the DR dynamics, which in the considered scenario propagates for just one step. 

A more interesting picture is provided by bank vulnerability values $V_\alpha$. As Figure \ref{fig:sect_reg_DR_1it}  shows, vulnerability now is not much driven by corporate loans shares ($C_\alpha/E_\alpha$), but correlates more with interbank loans shares ($A_\alpha/E_\alpha$). Indeed, the one-step shock propagation in the interbank layer, driven by $A_\alpha$, is much larger than the initial shock, determined (among other things) by $C_\alpha$. 
We therefore estimate the following minimal cross-sectional specification:
\begin{equation}
\log(V_\alpha) = \gamma
+ \beta_1 \log(E_\alpha)
+ \beta_2 (A_\alpha/E_\alpha)
+ \beta_3 (C_\alpha/E_\alpha)
+ \varepsilon_i
\label{eq:reg_vuln}
\end{equation}
where we used bank equity $E_\alpha$ as a proxy of size, while interbank assets $A_\alpha$ and corporate loans $C_\alpha$ are normalized to avoid collinearity and to represent leverage values. 
Regression results (Figure \ref{fig:sect_reg_DR_1it}) suggest that bank vulnerability is primarily driven by leverage rather than size. In the OLS specification, the model explains a substantial share of the variation in vulnerability ($R^2 = 0.63$), with both interbank exposure and corporate loan exposure being positive and statistically significant. By contrast, bank size, proxied by equity, is not statistically significant: balance sheet scale alone does not systematically determine vulnerability, once leverage ratios are accounted for.

The quantile regressions reveal important heterogeneity across the distribution of vulnerability. The coefficient of interbank share rises steadily from about 0.22 at the 10th percentile to more than 0.41 at the 90th percentile, indicating that interbank exposures play an increasingly dominant role among the most vulnerable banks. Corporate loan exposure also contributes positively across quantiles, though with a smaller and slightly declining effect at higher percentiles. In contrast, equity size shows significance only in the lowest quantile, where it enters with a negative coefficient, indicating that larger capital buffers may reduce vulnerability among the least fragile institutions, but this effect disappears for more vulnerable banks.

\section{Conclusions}

Understanding how shocks propagate between the real economy and the financial system is crucial for assessing systemic risk. Production networks and financial networks are deeply intertwined: disruptions affecting firms propagate through supply chains, reduce revenues, impair loan repayments, and ultimately transmit distress to banks -- which are amplified within the interbank market. However, empirical analyses of these mechanisms are typically constrained by the scarcity of detailed network data, since firm-to-firm transactions, credit exposures, and interbank relationships are largely confidential.

This work addresses this limitation by proposing a framework to reconstruct the multilayer structure linking firms and banks using only balance-sheet information. The approach combines reconstruction methods for production, bank–firm credit, and interbank networks into a unified multilayer representation of the economy. On this reconstructed structure, we simulate an ordered contagion mechanism in which shocks propagate from firms through supply chains, generate output losses measured by the Economic Systemic Risk Index (ESRI), translate into credit losses for banks captured by the Financial Systemic Risk Index (FSRI), and subsequently spread within the interbank market through equity deterioration computed by DebtRank (DR).

The illustrative application to the Italian economy highlights several important insights, which are in line with the empirical findings of the literature on network-based contagion. First, systemic importance is highly heterogeneous: only a very small fraction of firms generate large economic losses when they fail, while the vast majority have negligible systemic impact.  Second, the firms that are most critical for production are not necessarily the same as those posing the greatest risk to the banking system, indicating that economic and financial systemic importance overlap only partially. 
This highlights the importance of considering the full multilayer structure of the economy rather than focusing on a single network layer.
Finally, the empirical analysis of determinants of systemic impact suggests that financial contagion is strongly shaped by network exposures. Firms that are systemically important in the production network tend to generate larger losses for the banking sector, while banks’ vulnerability is largely determined by the composition of their assets, particularly their exposure to interbank lending and corporate loans. These results emphasize the role of network topology and cross-layer linkages in shaping systemic risk.
Furthermore, our framework can be used to simulate different scenarios of initial shocks, as well as to quantify their shot- and long-term impact. 
For instance, we showcase a scenario in which all firms within an industrial sector suffer small production losses, and study their immediate effects to the production and financial layers. 
In this way, we quantify the losses that occur before any mitigation action can possibly take place on the system. 

Overall, our framework demonstrates how multilayer reconstruction can enable network-based stress testing even in the absence of detailed microdata. 
From a policy perspective, the framework opens the possibility of constructing data-driven “digital twins” of economic systems that allow regulators to perform integrated stress tests across production and financial networks. Such tools could help identify firms whose disruption would generate disproportionate economic losses, detect banks whose balance-sheet composition makes them particularly exposed to real-economy shocks, and evaluate the systemic consequences of sectoral disruptions. At the same time, there are several directions for future research. Improving network reconstruction techniques, incorporating dynamic firm behavior and adaptive financial responses, and extending the framework to cross-border production and financial networks would allow a more realistic representation of shock propagation. 
Ultimately, by bridging production and financial networks within a unified multilayer representation, this work highlights how systemic risk emerges not only from individual institutions but from the architecture of the economic network itself, emphasizing the need for analytical tools capable of capturing the interconnected nature of modern economies.

\section*{Data and Code availability}

Codes to reconstruct the multilayer network and run the shock propagation dynamics using sample data are available at 
\url{https://github.com/mnlknt/bank-firm_multilayer_shocks}

\section*{Acknowledgements}

We acknowledge financial support from the National Recovery and Resilience Plan (NRRP), Mission 4 Component 2 Investment 1.1, funded by the European Union - NextGenerationEU: 
Call for tender No. 1409 of 14/09/2022 by the Italian Ministry of University and Research (MUR), Project Title: \emph{C2T - From Crises to Theory: towards a science of resilience and recovery for economic and financial systems}, Concession Decree No. 1381 of 01/09/2023, Project code P2022E93B8 - CUP E53D23018320001.
Call for tender No. 104 of 02/02/2022 by the Italian Ministry of University and Research (MUR), Project Title: \emph{RENet - Reconstructing economic networks: from physics to machine learning and back}, Concession Decree No. 957 of 30/06/2023, Project code 2022MTBB22 - CUP E53D23001770006;

	\bibliographystyle{apsrev4-2}
	
%

\newpage
\appendix

\renewcommand*{\thefigure}{A\arabic{figure}}
\renewcommand*{\thetable}{A\arabic{table}}
\setcounter{figure}{0}
\setcounter{table}{0}

\section*{Appendix}


    \begin{table}[h!]
		\centering
\begin{tabular}{lrrrrrrr}
\toprule
 & count & mean & median & std & min & max & skew \\
\midrule
Equity & 7109 & 77071 & 12761 & 923788 & 7 & 53644000 & 46 \\
Revenues & 7109 & 147411 & 17745 & 1721131 & 0 & 93717000 & 46 \\
Costs & 7109 & 109876 & 11801 & 1260132 & 0 & 69682000 & 44 \\
Bank debt & 7109 & 23644 & 2528 & 311761 & 0 & 21261000 & 50 \\
\bottomrule
\end{tabular}
\caption{Descriptive statistics for Italian firms dataset (2023). Monetary values in th EUR.}
		\label{tab:bankstats}
	\end{table}

	\begin{table}[h!]
		\centering
		\begin{tabular}{lrrrrrrr}
\toprule
 & count & mean & median & std & min & max & skew \\
\midrule
Total equity & 109 & 2418776 & 428691 & 8334136 & 110087 & 60303223 & 6 \\
Interbank assets & 109 & 3029873 & 196369 & 8503854 & 2 & 51380206 & 4 \\
Interbank liabilities & 109 & 5969979 & 816490 & 15413125 & 5 & 128435415 & 5 \\
Corporate loans & 109 & 6721815 & 1026702 & 23090709 & 2176 & 206761000 & 7 \\
\bottomrule
\end{tabular}
		\caption{Descriptive statistics for Italian banks dataset (2023). Monetary values in th EUR.}
		\label{tab:firmsstats}
	\end{table}

\begin{figure}[p]
	\centering
	\includegraphics[width=\textwidth]{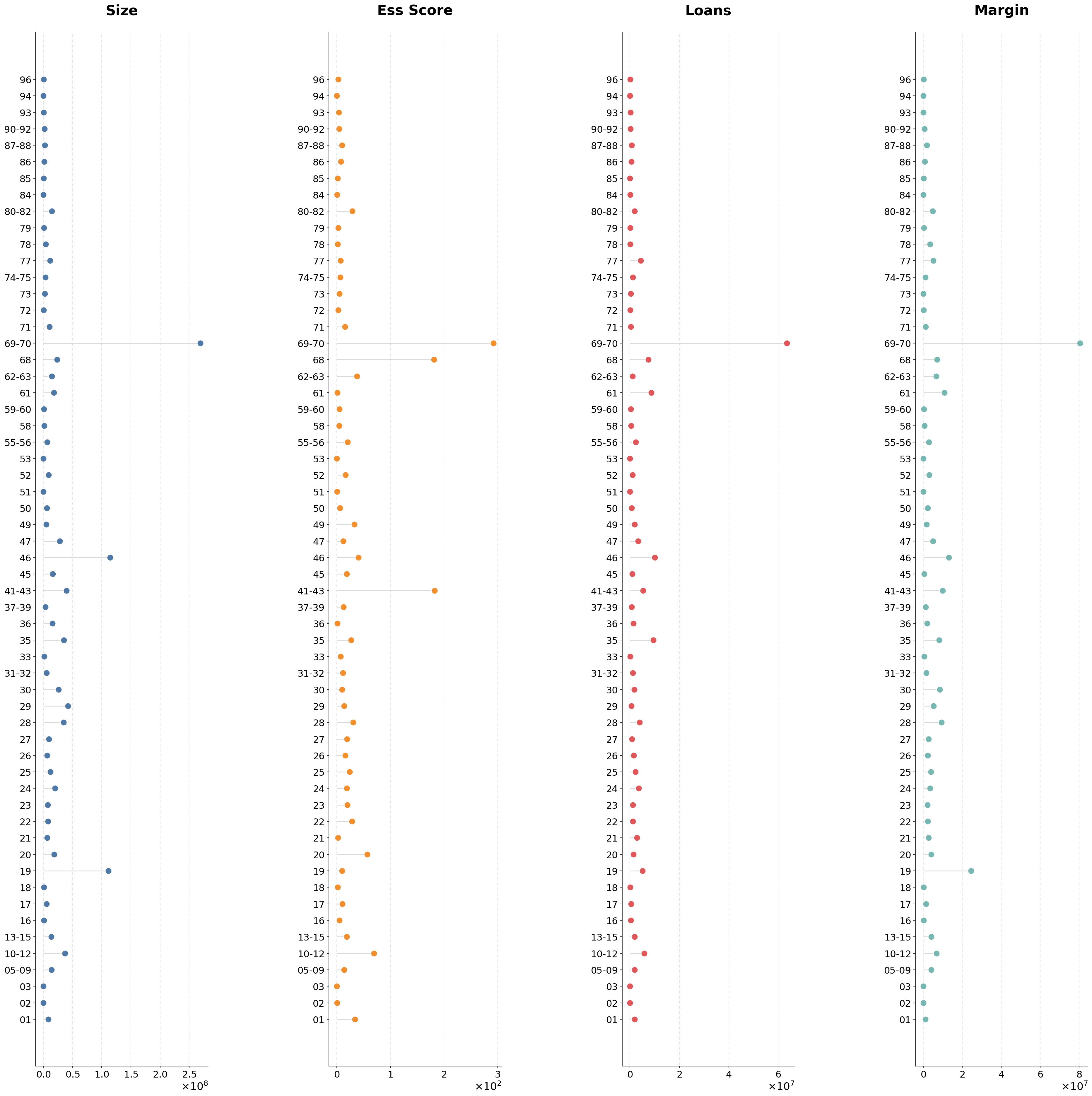}
    \caption{Balance sheet properties of the various NACE2 sectors.}
\label{fig:sector_rank}
\end{figure}

\end{document}